\documentclass[pdflatex,sn-mathphys-num,iicol]{sn-jnl}% Math and Physical Sciences Numbered Reference Style

\usepackage{graphicx}%
\usepackage{multirow}%
\usepackage{amsmath,amssymb,amsfonts}%
\usepackage{amsthm}%
\usepackage{mathrsfs}%
\usepackage[title]{appendix}%
\usepackage{xcolor}%
\usepackage{textcomp}%
\usepackage{manyfoot}%
\usepackage{booktabs}%
\usepackage{algorithm}%
\usepackage{algorithmicx}%
\usepackage{algpseudocode}%
\usepackage{listings}%
\usepackage{bm}

\usepackage{tikz}
\usetikzlibrary{positioning, shapes, arrows.meta, fit, calc}
\newcommand{\nodeWL}{6.5cm}   % node width
\newcommand{\nodeWR}{7.3cm}   % node width
\newcommand{\hsep}{1.5cm}    % horizontal separation between left and right groups
\newcommand{\firstarrow}{1.4cm}  % first arrow vertical spacing
\newcommand{\secondarrow}{2.2cm}  % second arrow vertical spacing
\tikzset{box/.style = {draw,rounded corners,align=center,minimum width=\nodeWL,inner sep=4pt, font=\footnotesize,fill=white},
	arrow/.style = {thick,-{Latex[length=1.2mm,width=1.2mm]}},
	bluebox/.style = {draw=blue!70!black,fill=blue!10,rounded corners,align=center,inner sep=2pt,font=\footnotesize}}

\def \d{\mathrm{d}}
\def \D{\mathrm{D}}

\theoremstyle{thmstyleone}%
\theoremstyle{thmstyletwo}%

\theoremstyle{thmstylethree}%

\begin{document}

\title{Coarse-graining particulate two-phase flow}

\author*[1]{\fnm{Thomas} \sur{P\"ahtz}}\email{0012136@zju.edu.cn}

\author[1]{\fnm{Yulan} \sur{Chen}}\email{una\_chen@zju.edu.cn}

\author*[1]{\fnm{Rui} \sur{Zhu}}\email{zhurui@zju.edu.cn}

\author[2]{\fnm{Katharina} \sur{Tholen}}\email{tholen@itp.uni-leipzig.de}

\author[1]{\fnm{Zhiguo} \sur{He}}\email{hezhiguo@zju.edu.cn}

\affil[1]{\orgdiv{Institute of Port, Coastal and Offshore Engineering, Ocean College}, \orgname{Zhejiang Univerity}, \orgaddress{\street{Zhoushan Campus}, \city{Zhoushan}, \postcode{316021}, \country{China}}}
\affil[2]{\orgdiv{Institute for Theoretical Physics}, \orgname{Leipzig University}, \orgaddress{\street{Br\"uderstra{\ss}e 16}, \city{Leipzig}, \postcode{04103}, \country{Germany}}}

\abstract{To acquire the ability to numerically study the rheology of particulate two-phase flows that lack scale separation, we present a general method to average or coarse-grain the equations of motion of a mixture of a continuous fluid of arbitrary rheology and non-Brownian particles, interacting via contacts, of arbitrary shapes and compositions. It universally covers ensemble and typical spatio-temporal averaging procedures and overcomes two shortcomings of existing methods. First, the derived micromechanical expressions for the coarse-grained fields are mathematically exact and formulated in a manner that allows a computationally cheap extraction from Direct Numerical Simulation-Discrete Element Method (DNS-DEM) simulations, avoiding the unlimited-order derivatives appearing in previous exact formulations. Second, the microscopic volume fraction of each particle is its corresponding indicator function, rather than the traditional volume-weighted delta distribution at its center of mass, to ensure that the resulting macroscopic fluid and solid volume fractions add precisely to unity. This leads to an additional contact stress contribution not seen in standard coarse-grained expressions for granular matter, and, for non-spherical particles, to particle-rotational contributions to translational solid phase balance equations. Many implementations of DNS-DEM simulations are based on Immersed Boundary Methods (IBMs), for which modifications of the coarse-graining method are necessary due to certain peculiarities of IBMs, such as the replacement of the particles' interiors by pseudo-fluid. We therefore derive mathematically exact adaptations of the coarse-graining method for two distinct common IBM versions, implement one version to obtain coarse-grained fields from sediment transport simulations based on this version, and validate the implementation.}




\maketitle

\section{Introduction} \label{Introduction}
In the context of granular matter, the term coarse-graining usually refers to the process of producing continuum balance equations from the equations of motion of a discrete system of non-Brownian particles that interact with one another via contacts~\citep{Goldhirsch10}. In a broader context, one may think of a mathematical procedure, usually some form of averaging, that smooths out small or microscopic scales, leading to a statistical or macroscopic description. As such, also Reynolds averaging of continuum mechanics balance equations~\citep{Germano92} may be considered coarse-graining, even though the fluid they model is already assumed continuous to begin with. Now, this study addresses the coarse-graining of a system of both fluid and particles, performed to obtain coupled macroscopic fluid and solid phase balance equations, where the coupling emerges from the flux through the fluid-solid interface of momentum, energy, and others.

Our main motivation behind developing a coarse-graining method for particulate two-phase flow is to acquire the ability to numerically study the rheology of systems that lack scale separation between the typical volume-equivalent particle radius $R$, or the typical separation distance between particles, and the macroscopic flow scale $L$. In other words, guided by small-scale simulations with particle-resolving numerical methods, such as the Direct Numerical Simulation (DNS) and Discrete Element Method (DEM), we aim to understand these systems' macroscopic behaviors in order to improve our ability to reliably model macroscopically heterogeneous fluid-particle flows at a large scale. A key example for a macroscopically highly heterogeneous system is fluid-driven sediment transport, where coarse-grained fluid and solid phase fields can change very rapidly over a fraction of $R$ in the direction normal to the sediment bed due to a typically sudden transition from a densely packed bed to a potentially very dilute transport layer~\citep{Fryetal24,Chassagneetal23,Tholenetal23}.

There have been numerous derivations of micromechanical expressions for the coarse-grained fields appearing in two-phase flow balance equations~\citep[e.g.,][]{AndersonJackson67,BuyevichShchelchkova79,Drew83,IshiiHibiki11,Croweetal12,Kolev15,Jackson97,ZhangProsperetti97,FintziPierson26}. However, most of them require scale separation, $R\ll L$, to be valid~\citep{AndersonJackson67,BuyevichShchelchkova79,Drew83,IshiiHibiki11,Croweetal12,Kolev15}, while those that do not are based on infinite-order Taylor expansions in $R/L$~\citep{Jackson97,ZhangProsperetti97,FintziPierson26}, implying that some of the coarse-grained fields are expressed as infinite sums of derivatives of ever higher order. Their extraction from DNS-DEM data is not only inconvenient but also computationally costly, since ever finer grids are required to reliably calculate these increasing-order derivatives.

When coarse-graining granular matter, one usually assigns each particle $p$ a microscopic volume fraction $\phi^p_\mathrm{mic}=V^p\delta[\bm{x}-\bm{x}^p]$ in space, $\bm{x}\in\mathbb{R}^3$, before averaging the equations of motion~\citep{Goldhirsch10}, where $V^p$ is the particle's volume, $\bm{x}^p$ its center of mass, and $\delta$ denotes the multivariate delta distribution. However, in the context of particulate two-phase flow, this definition is inappropriate, since it yields a macroscopic solid volume fraction $\langle\sum_p\phi^p_\mathrm{mic}\rangle$ that violates the desired condition $\langle\sum_p\phi^p_\mathrm{mic}\rangle+\beta_\mathrm{f}=1$~\citep{Jackson97}, where $\beta_\mathrm{f}$ is the macroscopic fluid volume fraction. Therefore, to satisfy this condition, one should instead use the alternative definition $\phi^p_\mathrm{mic}=X^p$, where the indicator function $X^p$ of a particle $p$ is equal to $1$ within its interior, $\bm{x}\in\mathbb{V}^p$, and equal to $0$ for $\bm{x}\notin\mathbb{V}^p$.

Here, we present a general method to coarse-grain particulate two-phase flow. It is mathematically exact for any $R/L$, its implementation is convenient and computationally cheap, and microscopic fields are based on $\phi^p_\mathrm{mic}=X^p$ to ensure macroscopic volume fraction consistency. The latter leads to an additional contact stress contribution not seen in standard coarse-grained expressions for granular matter, and, for non-spherical particles, to particle-rotational contributions to translational macroscopic solid phase balance equations. Furthermore, the method does not make assumptions about the fluid's rheology and the particles' shapes and compositions, permitting variations from particle to particle and mass density variations within each particle. Particles are also allowed to be mildly soft, meaning that their elastic deformations during contacts do not alter their overall shapes significantly and that the contact area of a contacting particle pair can be represented by a single contact point. Moreover, the method is based on a flexible averaging procedure that obeys simple rules which cover ensemble and typical spatio-temporal averages. Questions raised in previous studies regarding the coarse-graining scale required for uniqueness of the coarse-grained fields~\citep{Weinhartetal16} and the definition of a grid-independent particle fluctuation velocity~\citep{ArtoniRichard15b} are also addressed.

Many implementations of DNS-DEM simulations are based on the Immersed Boundary Method (IBM)~\citep[e.g.,][]{Uhlmann05,KempeFrohlich12b,Biegertetal17,Tschisgaleetal17,Zhuetal22,Bigotetal14,PiersonMagnaudet18,Fryetal24}, for which modifications of the coarse-graining method are necessary due to certain peculiarities of the IBM, such as the replacement of the particles' interiors by pseudo-fluid. We therefore derive two mathematically exact adaptations of the method that are applicable to the IBM after \citet{Uhlmann05} and the IBM after \citet{Bigotetal14}, respectively, and implement the former to obtain coarse-grained fields from IBM-based DNS-DEM simulations of steady, homogeneous sediment transport. To validate the implementation, we show that the extracted effective fluid and solid phase stresses are about equal to those obtained indirectly from integrating the respective momentum balances.

\section{Coarse-graining method} \label{CoarseGraining}
\subsection{Problem statement}
The goal is to obtain coupled macroscopic two-fluid balance equations for a fluid-particle mixture. The term balance equations can thereby refer to any kind of physical quantity, such as mass, momentum, energy, and others. Here, to ease the notation and illustrate the method's key principles, we derive the two-fluid mass and momentum balances for a mixture of an incompressible (but not necessarily Newtonian) fluid of constant density $\rho_\mathrm{f}$ and arbitrarily shaped particles of constant density $\rho_\mathrm{s}$, subjected to the constant gravitational acceleration $\bm{g}$. The most general case of an arbitrary balance equation, arbitrary density distributions of fluid, particles and their interiors, and arbitrary body force is presented in the supplementary material, which also includes more mathematical derivation details and additional information for the interested reader.

The microscopic problem can be summarized by the following fluid equations of motion:
\begin{align}
 \bm{\nabla}\cdot\bm{\tilde u}&=0, \label{MassFluidmic} \\
 \rho_\mathrm{f}(\partial_t\bm{\tilde u}+\bm{\tilde u}\cdot\bm{\nabla}\bm{\tilde u})&=\bm{\nabla}\cdot\bm{\sigma}+\rho_\mathrm{f}\bm{g}. \label{MomFluidmic}
\end{align}
They are valid within the fluid domain $\bm{x}\in\mathbb{V}_\mathrm{f}[t]$ and coupled with the following equations describing the motion of each particle $p$:
\begin{align}
 \rho_\mathrm{s}V^p\bm{\dot v}^p_\uparrow&=\int_{\mathbb{S}^p}\bm{n}^p\cdot\bm{\sigma}\d S+\rho_\mathrm{s}V^p\bm{g}+\bm{F}^{p}_\mathrm{c}, \label{MomParticle} \\
 \d_t\bm{I}^p\cdot\bm{\omega}^p&=\int_{\mathbb{S}^p}\bm{r}^p\times(\bm{n}^p\cdot\bm{\sigma})\d S+\bm{T}^{p}_\mathrm{c}. \label{AngularMomParticle}
\end{align}
In Eqs.~(\ref{MassFluidmic})-(\ref{AngularMomParticle}), $\bm{\tilde u}[\bm{x},t]$ is the fluid velocity, $\bm{\sigma}[\bm{x},t]$ the fluid stress tensor, $V^p$ is the constant volume of a particle $p$, $\bm{I}^p[t]$ its moment-of-inertia tensor, $\bm{v}^p_\uparrow[t]$ and $\bm{\omega}^p[t]$ its translational and angular velocities, respectively, $\mathbb{S}^p[t]=\partial\mathbb{V}^p[t]$ its surface domain, $\bm{n}^p$ the unit normal vector on its surface pointing outward, $\bm{r}^p[\bm{x},t]=\bm{x}-\bm{x}^p[t]$ the coordinate relative to its center of mass $\bm{x}^p[t]$, and $\bm{F}^{p}_\mathrm{c}[t]=\sum_q\bm{F}^{pq}_\mathrm{c}[t]$ and $\bm{T}^{p}_\mathrm{c}[t]=\sum_q\bm{T}^{pq}_\mathrm{c}[t]$ the contact force and torque, respectively, applied on it due to the interactions with other particles $q$ (by convention, $\bm{F}^{pp}_\mathrm{c}=0$, $\bm{T}^{pp}_\mathrm{c}=0$). Furthermore, the union $\mathbb{V}_\infty$ of the fluid ($\mathbb{V}_\mathrm{f}[t]$) and solid ($\mathbb{V}_\mathrm{s}[t]$) domains shall be a subset of $\mathbb{R}^3$: $\mathbb{V}_\mathrm{f}\cup\mathbb{V}_\mathrm{s}=\mathbb{V}_\infty\subseteq\mathbb{R}^3$, with $\mathbb{V}_\mathrm{s}=\cup_p\mathbb{V}^p$. The associated indicator functions therefore satisfy
\begin{equation}
 X_\infty=X_\mathrm{f}+X_\mathrm{s}=X_\mathrm{f}+\sum\nolimits_pX^p, \label{DefinitionChi}
\end{equation}
where $X_\infty[\bm{x}]$, $X_\mathrm{f}[\bm{x},t]$, $X_\mathrm{s}[\bm{x},t]$, and $X^p[\bm{x},t]$ are defined as being equal to $1$ in the interiors and equal to $0$ outside of $\mathbb{V}_\infty$, $\mathbb{V}_\mathrm{f}$, $\mathbb{V}_\mathrm{s}$, and $\mathbb{V}^p$, respectively.

\subsection{Generic averaging procedure} \label{Averaging}
\subsubsection{Averaging operator}
In the context of coarse-graining granular matter (indicated by the label `GM' on top of the equal sign below), microscopic fields $A[\bm{x},t]$ are usually averaged through a convolution with a kernel $\mathcal{W}[\bm{r},\tau]$ that integrates to unity~\citep{Babic97,Goldhirsch10}:
\begin{equation}
 \langle A[\bm{x},t]\rangle\stackrel{\text{GM}}{=}\int_{\mathbb{R}^4}A[\bm{x}-\bm{r},t-\tau]\mathcal{W}[\bm{r},\tau]\d^3r\d\tau. \label{SpatialAverage}
\end{equation}
In particular, when purely spatially averaging (i.e., $\mathcal{W}=\mathcal{W}_\ast[\bm{r}]\delta[\tau]$) the traditional microscopic volume fraction of a particle $p$, $\phi^p_\mathrm{mic}=V^p\delta[\bm{x}-\bm{x}^p]$, one simply obtains $V^p\mathcal{W}_\ast[\bm{x}-\bm{x}^p]$, which is why averaged fields in granular matter are usually expressed in terms of $\mathcal{W}_\ast[\bm{x}-\bm{x}^p]$ right from the start~\citep{Babic97,Goldhirsch10}. However, in our method to coarse-grain particulate two-phase flow, where microscopic fields are based on the indicator functions $X_\mathrm{f}$ and $X^p$, such as $\phi^p_\mathrm{mic}=X^p$ (see \S\ref{Introduction}), this simplification is no longer possible. Also, since we are interested in coarse-graining systems that lack scale separation, we cannot simply approximate $X^p$ by $V^p\delta[\bm{x}-\bm{x}^p]$. (The mathematically precise relationship between these two is given in Eq.~(\ref{XpDelta}).) All this means that, in contrast to the standard coarse-graining procedures for pure granular matter, the averaging operator $\langle\cdot\rangle$ will have to be carried along in the coming calculations. In addition, as we will see, the operator $\langle\cdot\rangle$ defined by Eq.~(\ref{SpatialAverage}) does generally not have all the properties we need for a mathematically exact derivation of the macroscopic balance equations.

To account for the above complications, we do generally not coarse-grain microscopic fields using Eq.~(\ref{SpatialAverage}), but instead express averaged fields using a generic averaging operator with the following properties:
\begin{alignat}{2}
 \langle cA_1+A_2\rangle&=c\langle A_1\rangle+\langle A_2\rangle,&\langle\bm{\nabla}A\rangle&=\bm{\nabla}\langle A\rangle, \nonumber \\
 \langle\partial_tA\rangle&=\partial_t\langle A\rangle,&\langle X_\infty A\rangle&=\langle A\rangle, \label{AveragingProcedure}
\end{alignat}
where $c$ is a constant. The first three properties, linearity and commutation with the spatial and temporal derivatives, are satisfied by the ensemble average~\citep{Zhangetal07b} and the spatio-temporal averaging procedures defined by Eq.~(\ref{SpatialAverage}). The fourth property ensures that averaging procedures only operate within $\mathbb{V}_\infty$, which is required for macroscopic volume fraction consistency as we will see. It also implies that external-boundary effects vanish due to
\begin{equation}
\begin{split}
 &\langle A\bm{\nabla}X_\infty\rangle=\langle\bm{\nabla}AX_\infty-X_\infty\bm{\nabla}A\rangle \\
 &=\bm{\nabla}\langle AX_\infty\rangle-\langle\bm{\nabla}A\rangle=\bm{\nabla}\langle A\rangle-\bm{\nabla}\langle A\rangle=0, \label{VanishingBoundary}
\end{split}
\end{equation}
which we will exploit in the derivation of the macroscopic fluid phase momentum balance. The fourth property is automatically satisfied for the ensemble and purely temporal averages and, if $\mathbb{V}_\infty=\mathbb{R}^3$, also for the spatio-temporal averages in Eq.~(\ref{SpatialAverage}). However, if $\mathbb{V}_\infty\subset\mathbb{R}^3$, it can be significantly violated for the latter averages. For example, kernels $\mathcal{W}$ that exhibit a compact support (i.e., $\mathcal{W}[\bm{r},\tau]=0$ for $|\bm{r}|>\zeta$) violate the fourth property at locations $\bm{x}$ that are separated from the boundary of $\mathbb{V}_\infty$ by less than the coarse-graining width $\zeta$, while kernels $\mathcal{W}$ that do not exhibit a compact support violate this property at almost all locations $\bm{x}$. At such rule-violating locations, some of the resulting derived macroscopic balance equations will not be identically valid. For this reason, we recommend to employ only those averaging procedures included in Eq.~(\ref{SpatialAverage}) that are based on kernels $\mathcal{W}$ with a compact support. The (typically few) rule-violating locations near the boundary of $\mathbb{V}_\infty$ can then be excluded when searching for rheological relationships. Note that the occurrence of rule-violating locations when using Eq.~(\ref{SpatialAverage}) cannot be prevented through a $\bm{x}$-dependent coarse-graining width $\zeta$ that becomes smaller near the boundary of $\mathbb{V}_\infty$, since this would lead to a violation of the commutation rule in Eq.~(\ref{AveragingProcedure}).

\subsubsection{Phase averages}
The fluid and solid phase averages of a field $A[\bm{x},t]$ are defined as weighted averages with respect to the microscopic concentrations $X_\mathrm{f}$ and $X_\mathrm{s}$, respectively:
\begin{align}
 &\langle A\rangle^\mathrm{f}\equiv\langle X_\mathrm{f}A\rangle/\beta_\mathrm{f},&&\langle A\rangle^\mathrm{s}\equiv\langle X_\mathrm{s}A\rangle/\beta_\mathrm{s}, \label{PhaseAverages}
\end{align}
where $\beta_\mathrm{f}\equiv\langle X_\mathrm{f}\rangle$ and $\beta_\mathrm{s}\equiv\langle X_\mathrm{s}\rangle$ are the macroscopic fluid and solid volume fractions, respectively. For the above definitions to make sense, $A$ must be well-defined for $\bm{x}\in\mathbb{V}_\mathrm{f}$ or $\bm{x}\in\mathbb{V}_\mathrm{s}$, respectively. Furthermore, two kinds of mixture averages of $A$, denoted as $\langle A\rangle$ and $\langle A\rangle_\rho$, are related to the fluid and solid phase averages through
\begin{align}
 \langle A\rangle&=\beta_\mathrm{f}\langle A\rangle^\mathrm{f}+\beta_\mathrm{s}\langle A\rangle^\mathrm{s}, \label{Mixture1} \\
 \langle A\rangle_\rho&\equiv\rho_\mathrm{m}^{-1}\left(\rho_\mathrm{f}\beta_\mathrm{f}\langle A\rangle^\mathrm{f}+\rho_\mathrm{s}\beta_\mathrm{s}\langle A\rangle^\mathrm{s}\right), \label{Mixture2}
\end{align}
where $\rho_\mathrm{m}\equiv\rho_\mathrm{f}\beta_\mathrm{f}+\rho_\mathrm{s}\beta_\mathrm{s}$ is the mixture density. Equation~(\ref{Mixture1}), which follows from $\langle A\rangle=\langle X_\infty A\rangle=\langle X_\mathrm{f}A\rangle+\langle X_\mathrm{s}A\rangle$, is one of the relations that requires the fourth rule in Eq.~(\ref{AveragingProcedure}) to be valid. In particular, the macroscopic volume fraction consistency condition, $\beta_\mathrm{f}+\beta_\mathrm{s}=1$, is a special case of Eq.~(\ref{Mixture1}).

\subsubsection{Reynolds rule, uniqueness, and coarse-graining scale} \label{ReynoldsRuleUniqueness}
The averaging rules in Eq.~(\ref{AveragingProcedure}) represent mathematical requirements of our coarse-graining method. If they are identically satisfied, the resulting balance equations are mathematically exact. However, to be also physically meaningful, the averaging operator should also satisfy the Reynolds rule, at least approximately:
\begin{equation}
 \langle\langle A_1\rangle A_2\rangle\stackrel{\mathrm{R}}{=}\langle A_1\rangle\langle A_2\rangle. \label{ReynoldsRule}
\end{equation}
Otherwise, like for filtering procedures used in Large Eddy Simulation~\citep{FurebyTabor97}, the resulting coarse-grained fields are not unique as they depend on the details of the averaging procedure. Excluding the trivial (and generally unphysical) cases of the identity operator and operators that average over all times and/or space, Eq.~(\ref{ReynoldsRule}) essentially states that the averaging should include the entire ensemble of microscopic states contributing to a macroscopic state at a given $\bm{x}$ and $t$, while not intermingling with different macroscopic states at surrounding $\bm{x}$ and $t$. There are only three kind of standard averaging procedures that identically satisfy this rule: the ensemble average, the infinite time average in steady systems, and infinite averages over a given spatial direction if a system is statistically homogeneous in that direction. Assuming that the ergodic hypothesis is true, all these three averages are equivalent, and averaged quantities are therefore unique. However, in practice, ensemble averaging is inconvenient and computationally costly, and systems are generally statistically heterogeneous in space and time. For this reason, one often resorts to sphere-symmetrical spatial averaging over a characteristic coarse-graining scale $\zeta$~\citep{AndersonJackson67,Weinhartetal16}. This scale should be sufficiently large to capture sufficiently many microscopic states, but not too large to substantially intermingle macroscopic states, which essentially necessitates scale separation between the important microscopic and macroscopic scales~\citep{AndersonJackson67}. In this regard, the important microscopic scales are often better represented by the typical separation distance between particles than by $R$~\citep{Goldhirsch08}, implying that, in systems with high solid volume fraction $\beta_\mathrm{s}$, such as dense granular flows, this condition might be satisfied even when scale separation between $R$ and $L$ is lacking. Then, there is often a goldilock zone, a range of $\zeta$ where sphere-symmetrically spatially averaged fields are almost unique~\citep{Weinhartetal16}, which is precisely the zone where such averages approximately satisfy the Reynolds rule.

\subsection{Derivation of macroscopic balance equations} \label{Derivation}
The derivation presented in this section follows a simple strategy. First, a balance equation is formulated at a microscopic, distributional level, involving generalized functions such as $X^p$ and $\delta[\bm{x}-\bm{x}^p]$. Afterward, a generic averaging operator $\langle\cdot\rangle$ that satisfies the rules in Eqs.~(\ref{AveragingProcedure}) and (\ref{ReynoldsRule}) is applied to both sides of the equation. The commutation rule then permits exchanging the order of differentiation and averaging, which leads to a macroscopic balance equation via exploiting the phase average definitions in Eqs.~(\ref{PhaseAverages})-(\ref{Mixture2}) and, if required, via carrying out a Reynolds decomposition.

\subsubsection{Global microscopic velocity fields}
In order to derive the two-fluid mass and momentum balances, we extend the microscopic fluid velocity $\bm{\tilde u}$, defined in $\mathbb{V}_\mathrm{f}$, to the entire fluid-particle domain $\mathbb{V}_\infty$ via equating it to the velocity of each particle $p$ within its respective domain $\mathbb{V}^p$:
\begin{equation}
 \bm{u}\equiv X_\mathrm{f}\bm{\tilde u}+\sum\nolimits_pX^p\left(\bm{v}^p_\uparrow+\bm{\omega}^p\times\bm{r}^p\right).
\end{equation}
Since the boundaries $\partial\mathbb{V}_\infty$ and $\partial\mathbb{V}_\mathrm{s}$ are (nearly) rigid, $\bm{\tilde u}$ exhibits impermeable boundary conditions, consistent with no-slip, Navier slip, and free-slip conditions. The interface-normal component of $\bm{u}$ is therefore continuous, implying the identities~\cite{Drew83}
\begin{equation}
 \D_tX_\mathrm{f}=\D_tX_\mathrm{s}=\D_tX^p=0, \label{DtX}
\end{equation}
where $\D_t\equiv\partial_t+\bm{u}\cdot\bm{\nabla}$ is the material derivative. In addition, since $\bm{\nabla}\cdot(\bm{v}^p_\uparrow+\bm{\omega}^p\times\bm{r}^p)=0$, also $\bm{\nabla}\cdot\bm{u}=0$ is satisfied. The identities in Eq.~(\ref{DtX}) are mathematically equivalent to Reynolds transport theorems~\cite{Drew83} for the fluid, solid, and particle material, respectively. They physically mean that a microscopic fluid, solid, or particle material point remains the same kind of material point when following its trajectory given by the velocity field $\bm{u}$.

In addition to $\bm{u}$, we also define the translational velocity field
\begin{equation}
 \bm{u}_\uparrow\equiv X_\mathrm{f}\bm{\tilde u}+\sum\nolimits_pX^p\bm{v}^p_\uparrow,
\end{equation}
which is the same as $\bm{u}$ but without the particle-rotational velocity contributions. Balance equations will be formulated in terms of $\bm{u}_\uparrow$ rather than $\bm{u}$ whenever possible.

\subsubsection{Macroscopic mass balance equations}
We multiply the microscopic mass balance $\bm{\nabla}\cdot\bm{u}=0$ with either $X_\mathrm{f}$ or $X_\mathrm{s}$, denoted as $X_\mathrm{f,s}$, use the chain rule of differentiation and Eq.~(\ref{DtX}), and obtain
\begin{equation}
 \partial_tX_\mathrm{f,s}+\bm{\nabla}\cdot X_\mathrm{f,s}\bm{u}=0.
\end{equation}
Averaging ($\langle\cdot\rangle$), using the rules and definitions in \S\ref{Averaging} and $X_\mathrm{f}\bm{u}=X_\mathrm{f}\bm{u}_\uparrow$, then yields
\begin{align}
 \partial_t\beta_\mathrm{f}+\bm{\nabla}\cdot\beta_\mathrm{f}\langle\bm{u}_\uparrow\rangle^\mathrm{f}&=0, \label{MassFluid} \\
 \partial_t\beta_\mathrm{s}+\bm{\nabla}\cdot\beta_\mathrm{s}\langle\bm{u}\rangle^\mathrm{s}&=0, \label{MassSolid} \\
 \bm{\nabla}\cdot\langle\bm{u}\rangle&=0, \label{VolumeMixture} \\
 \partial_t\rho_\mathrm{m}+\bm{\nabla}\cdot\rho_\mathrm{m}\langle\bm{u}\rangle_\rho&=0, \label{MassMixture}
\end{align}
where Eq.~(\ref{VolumeMixture}) follows from summing Eqs.~(\ref{MassFluid}) and (\ref{MassSolid}), and Eq.~(\ref{MassMixture}) from summing Eqs.~(\ref{MassFluid}) and (\ref{MassSolid}) after multiplication with $\rho_\mathrm{f}$ and $\rho_\mathrm{s}$, respectively. Equations~(\ref{MassFluid}), (\ref{MassSolid}), (\ref{VolumeMixture}), and (\ref{MassMixture}) represent the macroscopic fluid phase volume or mass conservation equation, the macroscopic solid phase volume or mass conservation equation, the macroscopic mixture volume conservation equation, and the macroscopic mixture mass conservation equation, respectively.

For sphere-symmetrical particles, the rotational particle velocity component is always normal to the particle surface, $\bm{n}^p\cdot(\bm{\omega}^p\times\bm{r}^p)=0$, which implies $\bm{u}\cdot\bm{\nabla}X_\mathrm{f,s}=\bm{u}_\uparrow\cdot\bm{\nabla}X_\mathrm{f,s}$~\citep{Drew83}. In this case, Eqs.~(\ref{MassSolid})-(\ref{MassMixture}) remain valid when formally replacing $\bm{u}$ by $\bm{u}_\uparrow$, which means that particle-rotational contributions vanish identically.

\subsubsection{Macroscopic momentum balance equations} \label{DerivationMomentum}
The derivation of the macroscopic momentum balance equations presented below is summarized in Fig.~\ref{SummaryDerivation}.
\begin{figure*}[t]
	\centering
	\begin{tikzpicture}[node distance=1cm and 1cm]
		\node[box] (L1) {
			\parbox{\nodeWL}{\centering \textbf{Microscopic fluid momentum balance ($\bm{x}\in \mathbb{V}_\mathrm{f}$)}\\ [4pt]
					$\rho_\mathrm{f}(\partial_t\tilde{\bm u}+\tilde{\bm u}\!\cdot\!\bm{\nabla}\tilde{\bm u})
					=\bm{\nabla}\!\cdot\!\bm{\sigma}+\rho_\mathrm{f}\bm{g}$}};
		\node[box,minimum height=0.5cm,below=\firstarrow of L1] (L2) {%定义第一个向下箭头长度
			\parbox{\nodeWL}{\centering \textbf{Microscopic fluid momentum balance ($\bm{x}\in \mathbb{V}_\infty$)}
				\begin{equation*}
					\begin{split}
						&\rho_\mathrm{f}\left(\partial_tX_\mathrm{f}\bm{u}_\uparrow+\bm{\nabla}\cdot X_\mathrm{f}\bm{u}_\uparrow\bm{u}_\uparrow\right)=\bm{\nabla}\cdot X_\mathrm{f}\bm{\sigma} \\
						&+\rho_\mathrm{f}X_\mathrm{f}\bm{g}-\bm{\sigma}\cdot\bm{\nabla}X_\infty+\sum\nolimits_p\bm{\sigma}\cdot\bm{\nabla}X^p
					\end{split}
		\end{equation*}\\[-6pt]}};
		\draw[arrow] (L1.south) -- (L2.north)
		node[pos=0.45,right,anchor=east,font=\footnotesize,align=right]{
			Multiply with $X_\mathrm{f}$,\\[1pt]
			use $\D_tX_\mathrm{f}=0$ and\\[1pt]
			$X_\mathrm{f}\bm{\tilde u}=X_\mathrm{f}\bm{u}_\uparrow$};
		\node[box, below=\secondarrow of L2] (L3) {\parbox{\nodeWL}{\centering \textbf{Macroscopic fluid phase momentum balance}\\[4pt]
				$\rho_\mathrm{f}\beta_\mathrm{f}\D^\mathrm{f}_t\langle\bm{u}_\uparrow\rangle^\mathrm{f}=\bm{\nabla}\cdot\bm{\sigma^\mathrm{f}}+\rho_\mathrm{f}\beta_\mathrm{f}\bm{g}-\beta_\mathrm{s}\langle\bm{f}_\mathrm{h}\rangle^\mathrm{s}$}};
		\draw[arrow] (L2.south) -- (L3.north)
		node[pos=0.65,left,anchor=east,font=\footnotesize,align=right]{Averaging: $\langle\cdot\rangle$ \\Averaging rules\\Phase average definitions\\Reynolds decomposition};

		\node[bluebox,minimum height=1.25cm,right=0.7cm of $(L2)!0.42!(L3)$] (Lside) {
			\parbox{2.5cm}{Express $\bm{\sigma}\!\cdot\!\bm{\nabla} X^p$ as \\[4pt]
				$-X^p\bm{f}^p_\mathrm{h}+\bm{\nabla}\!\cdot(\dots)$,\\[2pt]
				 as shown in Eq.~(\ref{InteractionTerm2})}};
		\coordinate (midL23) at ($(L2.south)!0.3!(L3.north)$);
		\coordinate (startPoint) at ($(Lside.north west)!0.5!(Lside.south west)$);
		\draw[arrow] (startPoint) -- (midL23)
		node[pos=0.5,above,anchor=west,align=left,font=\scriptsize]{};

		%%% === RIGHT SIDE ===
		\node[box, right=\hsep of L1] (R1) {
			\parbox{\nodeWR}{\centering \textbf{Distributional Reynolds transport theorem} ($\bm{x}\in \mathbb{V}_\infty$) \\[4pt]
					$\D_tX^p=0$}};
		\node[box, below=\firstarrow of R1] (R2) {\parbox{\nodeWR}{\centering \textbf{Microscopic particle momentum balance} ($\bm{x}\in \mathbb{V}_\infty$) \\[4pt]
			$\rho_\mathrm{s}\left(\partial_tX^p\bm{u}_\uparrow+\bm{\nabla}\cdot X^p\bm{u}\bm{u}_\uparrow\right)=X^p\left(\rho_\mathrm{s}\bm{g}+\bm{f}^p_\mathrm{h}+\bm{f}^p_\mathrm{c}\right)$}};
		\draw[arrow] (R1.south) -- (R2.north)
		node[pos=0.5,right,anchor=east,align=right,font=\footnotesize]{Multiply with $ \rho_\mathrm{s}\bm{u}_\uparrow$, use $\bm{\nabla}\cdot\bm{u}=0$,\\[1pt]
		$\bm{u}_\uparrow\D_tX^p=\D_tX^p\bm{u}_\uparrow -X^p\D_t\bm{u}_\uparrow$, \\[1pt]
		and $\rho_\mathrm{s}X^p\D_t\bm{u}_\uparrow=X^p(\rho_\mathrm{s}\bm{g}+\bm{f}^p_\mathrm{h}+\bm{f}^p_\mathrm{c})$};
		\node[box, below=2.8cm of R2] (R3) {\parbox{\nodeWR}{\centering \textbf{Macroscopic solid phase momentum balance}\\[4pt]
	 $\rho_\mathrm{s}\beta_\mathrm{s}\D^\mathrm{s}_t\langle\bm{u}_\uparrow\rangle^\mathrm{s}=\bm{\nabla}\cdot\bm{\sigma^\mathrm{s}}+\rho_\mathrm{s}\beta_\mathrm{s}\bm{g}+\beta_\mathrm{s}\langle\bm{f}_\mathrm{h}\rangle^\mathrm{s}$}};
		\draw[arrow] (R2.south) -- (R3.north)
		node[pos=0.7,left,anchor=east,font=\footnotesize,align=right]{Summing, averaging: $\langle\sum_p\cdot\rangle$ \\Averaging rules\\Phase average definitions\\Reynolds decomposition};

		%%% === BLUE SIDE BOX ===
	\node[bluebox,minimum height=0.9cm,right=0.4cm of $(R2)!0.35!(R3)$] (Rside) {\parbox{3.2cm}{Express $X^p\bm{f}^p_\mathrm{c}$ as	$\bm{\nabla}\cdot(\dots)$,\\[2pt]
		as shown in  Eqs.~(\ref{Contact1})-(\ref{Contact2})}};

	%%% === BLUE BOX → vertical arrow center ===
	\coordinate (midL23) at ($(R2.south)!0.295!(R3.north)$);
	\coordinate (startPoint) at ($(Rside.north west)!0.5!(Rside.south west)$);
	\draw[arrow] (startPoint) -- (midL23)
	node[pos=0.5,above,anchor=west,align=left,font=\scriptsize]{	};
	\end{tikzpicture}
	\caption{Summary of the derivation of the macroscopic momentum balance equations. The bluish boxes mark the most important derivation steps. They are responsible for the novel elements in the final expressions for the effective fluid phase ($\bm{\sigma^\mathrm{f}}$) and solid phase ($\bm{\sigma^\mathrm{s}}$) stress tensors when compared with expressions from previous studies.}
	\label{SummaryDerivation}
\end{figure*}

We multiply the microscopic fluid momentum balance, Eq.~(\ref{MomFluidmic}), with $X_\mathrm{f}$, use $X_\mathrm{f}\bm{\tilde u}=X_\mathrm{f}\bm{u}_\uparrow$, the chain rule of differentiation, and Eqs.~(\ref{DefinitionChi}) and (\ref{DtX}), and obtain
\begin{equation}
\begin{split}
 &\rho_\mathrm{f}\left(\partial_tX_\mathrm{f}\bm{u}_\uparrow+\bm{\nabla}\cdot X_\mathrm{f}\bm{u}_\uparrow\bm{u}_\uparrow\right)=\bm{\nabla}\cdot X_\mathrm{f}\bm{\sigma} \\
 &+\rho_\mathrm{f}X_\mathrm{f}\bm{g}-\bm{\sigma}\cdot\bm{\nabla}X_\infty+\sum\nolimits_p\bm{\sigma}\cdot\bm{\nabla}X^p.
\end{split}
\end{equation}
Averaging ($\langle\cdot\rangle$), employing the rules and definitions in \S\ref{Averaging}, then yields
\begin{equation}
\begin{split}
 &\rho_\mathrm{f}\left(\partial_t\beta_\mathrm{f}\langle\bm{u}_\uparrow\rangle^\mathrm{f}+\bm{\nabla}\cdot\beta_\mathrm{f}\langle\bm{u}_\uparrow\bm{u}_\uparrow\rangle^\mathrm{f}\right) \\
 &=\bm{\nabla}\cdot\beta_\mathrm{f}\langle\bm{\sigma}\rangle^\mathrm{f}+\rho_\mathrm{f}\beta_\mathrm{f}\bm{g}+\sum\nolimits_p\langle\bm{\sigma}\cdot\bm{\nabla}X^p\rangle.
\end{split} \label{MomFluidhelp}
\end{equation}
In particular, we used Eq.~(\ref{VanishingBoundary}) to conclude $\langle\bm{\sigma}\cdot\bm{\nabla}X_\infty\rangle=0$. Carrying out a Reynolds decomposition on the left-hand side of Eq.~(\ref{MomFluidhelp}), using Eq.~(\ref{MassFluid}), Eq.~(\ref{MomFluidhelp}) becomes
\begin{equation}
\begin{split}
 \rho_\mathrm{f}\beta_\mathrm{f}\D^\mathrm{f}_t\langle\bm{u}_\uparrow\rangle^\mathrm{f}=&\;\bm{\nabla}\cdot\left(\beta_\mathrm{f}\langle\bm{\sigma}\rangle^\mathrm{f}+\bm{\sigma^\mathrm{f}_\mathrm{Re}}\right)+\rho_\mathrm{f}\beta_\mathrm{f}\bm{g} \\
 &+\sum\nolimits_p\langle\bm{\sigma}\cdot\bm{\nabla}X^p\rangle,
\end{split} \label{MomFluidhelp2}
\end{equation}
where $\D^\mathrm{f}_t\equiv\partial_t+\langle\bm{u}_\uparrow\rangle^\mathrm{f}\cdot\bm{\nabla}$ is the associated material derivative and
\begin{equation}
\begin{split}
 \bm{\sigma^\mathrm{f}_\mathrm{Re}}&\equiv-\rho_\mathrm{f}\beta_\mathrm{f}\left(\langle\bm{u}_\uparrow\bm{u}_\uparrow\rangle^\mathrm{f}-\langle\bm{u}_\uparrow\rangle^\mathrm{f}\langle\bm{u}_\uparrow\rangle^\mathrm{f}\right) \\
 &\stackrel{\mathrm{R}}{=}-\rho_\mathrm{f}\beta_\mathrm{f}\left\langle\left(\bm{u}_\uparrow-\langle\bm{u}_\uparrow\rangle^\mathrm{f}\right)\left(\bm{u}_\uparrow-\langle\bm{u}_\uparrow\rangle^\mathrm{f}\right)\right\rangle^\mathrm{f}
\end{split} \label{FluidReynolds}
\end{equation}
the fluid phase Reynolds stress. As indicated by the symbol `$\stackrel{\mathrm{R}}{=}$', the equality on the bottom line requires the Reynolds rule, Eq.~(\ref{ReynoldsRule}), to be valid.

Existing derivations of the macroscopic fluid phase momentum balance equation essentially agree until Eq.~(\ref{MomFluidhelp2}) is reached~\citep{AndersonJackson67,BuyevichShchelchkova79,Drew83,IshiiHibiki11,Croweetal12,Kolev15,Jackson97,ZhangProsperetti97,FintziPierson26}. Then, however, they start to diverge from one another in the mathematical treatment of the term $\sum\nolimits_p\langle\bm{\sigma}\cdot\bm{\nabla}X^p\rangle$, which encodes fluid-solid interactions as $\bm{\nabla}X^p$ is nonzero only at the fluid-solid interface. In Appendix~\ref{Identities}, we show that $\bm{\sigma}\cdot\bm{\nabla}X^p$ can be expressed as
\begin{equation}
\begin{split}
 &\bm{\sigma}\cdot\bm{\nabla}X^p \\
 =&-\left(X^p+\bm{\nabla}\cdot\int_{\mathbb{V}^p}\left(\bm{r}^p\delta^p_{\mathrm{L}\bm{x}}\right)[\bm{y}]\d V_y\right)\bm{f}^p_\mathrm{h} \\
 &+\bm{\nabla}\cdot\int_{\mathbb{S}^p}\left(\bm{r}^p\bm{n}^p\cdot\bm{\sigma}\delta^p_{\mathrm{L}\bm{x}}\right)[\bm{y}]\d S_y,
\end{split} \label{InteractionTerm}
\end{equation}
where $\bm{f}^p_\mathrm{h}=\bm{F}^p_\mathrm{h}/V^p$ is the hydrodynamic force acting on a particle $p$ per unit particle volume $V^p$, with $\bm{F}^p_\mathrm{h}$ given by
\begin{equation}
 \bm{F}^p_\mathrm{h}=\int_{\mathbb{S}^p}\bm{n}^p\cdot\bm{\sigma}\d S,
\end{equation}
and $\delta^p_{\mathrm{L}\bm{x}}[\bm{y}]$ denotes a `delta line' connecting its center of mass $\bm{x}^p$ with $\bm{y}$:
\begin{equation}
 \delta^p_{\mathrm{L}\bm{x}}[\bm{y}]\equiv\int_0^1\delta[\bm{x}-\bm{x}^p-\lambda(\bm{y}-\bm{x}^p)]\d\lambda. \label{DeltaLine}
\end{equation}
To write Eq.~(\ref{InteractionTerm}) in a more compact form, we introduce the notations $\underline{A}_{\mathbb{V}^p}\equiv\frac{1}{V^p}\int_{\mathbb{V}^p}A\d V$ and $\underline{A}_{\mathbb{S}^p}\equiv\frac{1}{S^p}\int_{\mathbb{S}^p}A\d S$, with $S^p$ the surface area of a particle $p$, yielding
\begin{equation}
\begin{split}
 \bm{\sigma}\cdot\bm{\nabla}X^p=&-X^p\bm{f}^p_\mathrm{h}-\bm{\nabla}\cdot\underline{\bm{r}^p\delta^p_{\mathrm{L}\bm{x}}}_{\mathbb{V}^p}\bm{F}^p_\mathrm{h} \\
 &+\bm{\nabla}\cdot S^p\underline{\bm{r}^p\bm{n}^p\cdot\bm{\sigma}\delta^p_{\mathrm{L}\bm{x}}}_{\mathbb{S}^p}.
\end{split} \label{InteractionTerm2}
\end{equation}
Note that $\bm{\nabla}$ in both Eqs.~(\ref{InteractionTerm}) and (\ref{InteractionTerm2}) acts on the `$\bm{x}$' within $\delta^p_{\mathrm{L}\bm{x}}$.

Equation~(\ref{InteractionTerm2}) expresses the fluid-solid interaction term $\bm{\sigma}\cdot\bm{\nabla}X^p$ as the sum of the negative microscopic hydrodynamic force per unit volume, $-X^p\bm{f}^p_\mathrm{h}$, and a divergence term. Upon summing over all particles and averaging, since $\langle\cdot\rangle$ commutes with $\bm{\nabla}$, we obtain the macroscopic fluid phase momentum balance, in which the latter term becomes the divergence of a stress associated with fluid-solid interactions, while the former term becomes the negative macroscopic hydrodynamic force density:
\begin{equation}
 \rho_\mathrm{f}\beta_\mathrm{f}\D^\mathrm{f}_t\langle\bm{u}_\uparrow\rangle^\mathrm{f}=\bm{\nabla}\cdot\bm{\sigma^\mathrm{f}}+\rho_\mathrm{f}\beta_\mathrm{f}\bm{g}-\beta_\mathrm{s}\langle\bm{f}_\mathrm{h}\rangle^\mathrm{s}, \label{MomFluid}
\end{equation}
where $\bm{f}_\mathrm{h}\equiv\sum_pX^p\bm{f}^p_\mathrm{h}$ is the microscopic hydrodynamic force density and $\bm{\sigma^\mathrm{f}}$ the effective fluid phase stress tensor, given by
\begin{align}
 \bm{\sigma^\mathrm{f}}&\equiv\beta_\mathrm{f}\langle\bm{\sigma}\rangle^\mathrm{f}+\bm{\sigma^\mathrm{f}_\mathrm{Re}}+\bm{\sigma^\mathrm{f}_\mathrm{s}}, \label{SigmaFluid} \\
 \bm{\sigma^\mathrm{f}_\mathrm{s}}&\equiv\left\langle\sum\nolimits_p\left(S^p\underline{\bm{r}^p\bm{n}^p\cdot\bm{\sigma}\delta^p_{\mathrm{L}\bm{x}}}_{\mathbb{S}^p}-\underline{\bm{r}^p\delta^p_{\mathrm{L}\bm{x}}}_{\mathbb{V}^p}\bm{F}^p_\mathrm{h}\right)\right\rangle.
\end{align}

To derive the macroscopic solid phase momentum balance, we multiply $\D_tX^p=0$, Eq.~(\ref{DtX}), with $\rho_\mathrm{s}\bm{u}_\uparrow$, use $\bm{\nabla}\cdot\bm{u}=0$, the chain rule of differentiation, and $\rho_\mathrm{s}X^p\D_t\bm{u}_\uparrow=X^p\rho_\mathrm{s}\bm{\dot v}^p_\uparrow=X^p(\rho_\mathrm{s}\bm{g}+\bm{f}^p_\mathrm{h}+\bm{f}^p_\mathrm{c})$, and obtain
\begin{equation}
 \rho_\mathrm{s}\left(\partial_tX^p\bm{u}_\uparrow+\bm{\nabla}\cdot X^p\bm{u}\bm{u}_\uparrow\right)=X^p\left(\rho_\mathrm{s}\bm{g}+\bm{f}^p_\mathrm{h}+\bm{f}^p_\mathrm{c}\right),
\end{equation}
where $\bm{f}^p_\mathrm{c}\equiv\bm{F}^p_\mathrm{c}/V^p$ is the contact force on a particle $p$ per unit particle volume $V^p$. Summing over all particles and averaging ($\langle\cdot\rangle$), employing the rules and definitions in \S\ref{Averaging}, then yields
\begin{equation}
\begin{split}
 &\rho_\mathrm{s}\left(\partial_t\beta_\mathrm{s}\langle\bm{u}_\uparrow\rangle^\mathrm{s}+\bm{\nabla}\cdot\beta_\mathrm{s}\langle\bm{u}\bm{u}_\uparrow\rangle^\mathrm{s}\right) \\
 &=\rho_\mathrm{s}\beta_\mathrm{s}\bm{g}+\beta_\mathrm{s}\langle\bm{f}_\mathrm{h}\rangle^\mathrm{s}+\left\langle\sum\nolimits_pX^p\bm{f}^p_\mathrm{c}\right\rangle.
\end{split} \label{MomSolidhelp}
\end{equation}

As shown in Appendix~\ref{Identities}, the contact force term in Eq.~(\ref{MomSolidhelp}) can be manipulated via
\begin{equation}
 X^p\bm{f}^p_\mathrm{c}=\bm{F}^p_\mathrm{c}\delta[\bm{x}-\bm{x}^p]-\bm{\nabla}\cdot\underline{\bm{r}^p\delta^p_{\mathrm{L}\bm{x}}}_{\mathbb{V}^p}\bm{F}^p_\mathrm{c}, \label{Contact1}
\end{equation}
where the first term on the right-hand side, after summing over all particles, can be transformed into a divergence in the standard manner using $\bm{F}^{pq}_\mathrm{c}=-\bm{F}^{qp}_\mathrm{c}$ and Eq.~(\ref{DeltaId})~\citep{Weinhartetal12a,Weinhartetal16}:
\begin{equation}
\begin{split}
 &\sum\nolimits_p\bm{F}^p_\mathrm{c}\delta[\bm{x}-\bm{x}^p]=\sum\nolimits_{pq}\bm{F}^{pq}_\mathrm{c}\delta[\bm{x}-\bm{x}^p] \\
 &=\frac{1}{2}\sum\nolimits_{pq}\bm{F}^{pq}_\mathrm{c}\left(\delta[\bm{x}-\bm{x}^p]-\delta[\bm{x}-\bm{x}^q]\right) \\
 &=\bm{\nabla}\cdot\frac{1}{2}\sum\nolimits_{pq}\left(\bm{r}^{pq}_\mathrm{c}\delta^{pq}_{\mathrm{cL}\bm{x}}-\bm{r}^{qp}_\mathrm{c}\delta^{qp}_{\mathrm{cL}\bm{x}}\right)\bm{F}^{pq}_\mathrm{c}.
\end{split} \label{Contact2}
\end{equation}
It is based on two delta lines associated with a pair of contacting particles $p$ and $q$: $\delta^{pq}_{\mathrm{cL}\bm{x}}\equiv\delta^p_{\mathrm{L}\bm{x}}[\bm{x}^{pq}_\mathrm{c}]$, connecting $\bm{x}^p$ with the contact point $\bm{x}^{pq}_\mathrm{c}=\bm{x}^{qp}_\mathrm{c}$, and $\delta^{pq}_{\mathrm{cL}\bm{x}}\equiv\delta^q_{\mathrm{L}\bm{x}}[\bm{x}^{pq}_\mathrm{c}]$, connecting $\bm{x}^q$ with $\bm{x}^{pq}_\mathrm{c}$, while $\bm{r}^{pq}_\mathrm{c}\equiv\bm{x}^{pq}_\mathrm{c}-\bm{x}^p$ and $\bm{r}^{qp}_\mathrm{c}\equiv\bm{x}^{pq}_\mathrm{c}-\bm{x}^q$.

Inserting Eqs.~(\ref{Contact1}) and (\ref{Contact2}) in Eq.~(\ref{MomSolidhelp}) and carrying out a Reynolds decomposition analogous to that for the macroscopic fluid phase momentum balance, using Eq.~(\ref{MassSolid}), we obtain
\begin{equation}
 \rho_\mathrm{s}\beta_\mathrm{s}\D^\mathrm{s}_t\langle\bm{u}_\uparrow\rangle^\mathrm{s}=\bm{\nabla}\cdot\bm{\sigma^\mathrm{s}}+\rho_\mathrm{s}\beta_\mathrm{s}\bm{g}+\beta_\mathrm{s}\langle\bm{f}_\mathrm{h}\rangle^\mathrm{s}, \label{MomSolid}
\end{equation}
where $\D^\mathrm{s}_t\equiv\partial_t+\langle\bm{u}\rangle^\mathrm{s}\cdot\bm{\nabla}$ is the associated material derivative and $\bm{\sigma^\mathrm{s}}$ the effective solid phase stress tensor, given by
\begin{align}
 \bm{\sigma^\mathrm{s}}\equiv&\;\bm{\sigma^\mathrm{s}_\mathrm{Re}}+\bm{\sigma^\mathrm{s}_\mathrm{c}}, \label{SigmaSolid} \\
\begin{split}
 \bm{\sigma^\mathrm{s}_\mathrm{Re}}\equiv&-\rho_\mathrm{s}\beta_\mathrm{s}\left(\langle\bm{u}\bm{u}_\uparrow\rangle^\mathrm{s}-\langle\bm{u}\rangle^\mathrm{s}\langle\bm{u}_\uparrow\rangle^\mathrm{s}\right) \\
 \stackrel{\mathrm{R}}{=}&-\rho_\mathrm{s}\beta_\mathrm{s}\left\langle\left(\bm{u}-\langle\bm{u}\rangle^\mathrm{s}\right)\left(\bm{u}_\uparrow-\langle\bm{u}_\uparrow\rangle^\mathrm{s}\right)\right\rangle^\mathrm{s},
\end{split} \label{SigmaSolidReynolds} \\
\begin{split}
 \bm{\sigma^\mathrm{s}_\mathrm{c}}\equiv&\;\left\langle\frac{1}{2}\sum\nolimits_{pq}\left(\bm{r}^{pq}_\mathrm{c}\delta^{pq}_{\mathrm{cL}\bm{x}}-\bm{r}^{qp}_\mathrm{c}\delta^{qp}_{\mathrm{cL}\bm{x}}\right)\bm{F}^{pq}_\mathrm{c}\right\rangle \\
 &-\left\langle\sum\nolimits_p\underline{\bm{r}^p\delta^p_{\mathrm{L}\bm{x}}}_{\mathbb{V}^p}\bm{F}^p_\mathrm{c}\right\rangle.
\end{split} \label{SigmaSolidContact}
\end{align}
It consists of the solid phase Reynolds stress $\bm{\sigma^\mathrm{s}_\mathrm{Re}}$ and the contact stress $\bm{\sigma^\mathrm{s}_\mathrm{c}}$. In contrast to standard expressions for granular matter~\citep{Babic97,Goldhirsch10,Weinhartetal12a,Weinhartetal16}, $\bm{\sigma^\mathrm{s}_\mathrm{Re}}$ contains particle-rotational contributions, and $\bm{\sigma^\mathrm{s}_\mathrm{c}}$ the additional term $-\langle\sum_p\underline{\bm{r}^p\delta^p_{\mathrm{L}\bm{x}}}_{\mathbb{V}^p}\bm{F}^p_\mathrm{c}\rangle$.

Similar to before, in the case of sphere-symmetrical particles, the particle-rotational contributions to $\D^\mathrm{s}_t\langle\bm{u}_\uparrow\rangle^\mathrm{s}$ and $\bm{\sigma^\mathrm{s}_\mathrm{Re}}$ cancel each other out identically due to $\bm{n}^p\cdot(\bm{\omega}^p\times\bm{r}^p)=0$, and $\bm{u}$ can therefore be \textit{simultaneously} formally replaced by $\bm{u}_\uparrow$ in the definitions of $\D^\mathrm{s}_t$ and $\bm{\sigma^\mathrm{s}_\mathrm{Re}}$. In addition, the term $\bm{r}^{pq}_\mathrm{c}\delta^{pq}_{\mathrm{cL}\bm{x}}-\bm{r}^{qp}_\mathrm{c}\delta^{qp}_{\mathrm{cL}\bm{x}}$ within the contact stress $\bm{\sigma^\mathrm{s}_\mathrm{c}}$ simplifies to its standard form $(\bm{x}^q-\bm{x}^p)\delta^p_{\mathrm{L}\bm{x}}[\bm{x}^q]$~\citep{Babic97,Goldhirsch10}.

Summing the macroscopic fluid phase and solid phase momentum balances, Eqs.~(\ref{MomFluid}) and (\ref{MomSolid}), leads to the macroscopic mixture momentum balance
\begin{equation}
 \rho_\mathrm{m}\D^\mathrm{m}_t\langle\bm{u}_\uparrow\rangle_\rho=\bm{\nabla}\cdot\bm{\sigma^\mathrm{m}}+\rho_\mathrm{m}\bm{g},
\end{equation}
where $\D^\mathrm{m}_t\equiv\partial_t+\langle\bm{u}\rangle_\rho\cdot\bm{\nabla}$ is the associated material derivative and $\bm{\sigma^\mathrm{m}}$ the effective mixture stress tensor, given by
\begin{align}
 \bm{\sigma^\mathrm{m}}&\equiv\beta_\mathrm{f}\langle\bm{\sigma}\rangle^\mathrm{f}+\bm{\sigma^\mathrm{f}_\mathrm{s}}+\bm{\sigma^\mathrm{s}_\mathrm{c}}+\bm{\sigma^\mathrm{m}_\mathrm{Re}}, \label{SigmaMixture} \\
\begin{split}
 \bm{\sigma^\mathrm{m}_\mathrm{Re}}&\equiv-\rho_\mathrm{m}\left(\langle\bm{u}\bm{u}_\uparrow\rangle_\rho-\langle\bm{u}\rangle_\rho\langle\bm{u}_\uparrow\rangle_\rho\right) \\
 &\stackrel{\mathrm{R}}{=}-\rho_\mathrm{m}\left\langle\left(\bm{u}-\langle\bm{u}\rangle_\rho\right)\left(\bm{u}_\uparrow-\langle\bm{u}_\uparrow\rangle_\rho\right)\right\rangle_\rho,
\end{split} \label{MixtureReynolds}
\end{align}
with $\bm{\sigma^\mathrm{m}_\mathrm{Re}}$ the mixture Reynolds stress.

\subsubsection{Note on Reynolds stresses and fluctuation velocities}
When numerical implementing the coarse-graining method, one should use the top-line expressions in Eqs.~(\ref{FluidReynolds}), (\ref{SigmaSolidReynolds}), and (\ref{MixtureReynolds}) to calculate the Reynolds stresses rather than the usual respective bottom-line expressions based on the fluctuation velocities, since the latter require the Reynolds rule to be valid, which is almost never identically satisfied in practice. In fact, in the fluid mechanics community, the analog of the top line of Eq.~(\ref{FluidReynolds}) has been used for several decades to calculate the pure-fluid Reynolds stress tensor in numerical simulations~\citep{Germano92}.

\citet{ArtoniRichard15b} introduced an alternative definition of the fluctuation velocity for granular flows that eliminates the grid dependence of the solid phase Reynolds stress (based on the traditional microscopic volume fraction of a particle $p$, $\phi^p_\mathrm{mic}=V^p\delta[\bm{x}-\bm{x}^p]$) for systems with slowly-varying shear rate. However, since our coarse-graining method requires the Reynolds rule to be at least approximately satisfied, which implies almost no grid dependence of the coarse-grained fields (\S\ref{ReynoldsRuleUniqueness}), there is no good justification for deviating from using Eq.~(\ref{SigmaSolidReynolds}) to calculate $\bm{\sigma^\mathrm{s}_\mathrm{Re}}$.

\section{Coarse-graining IBM-based DNS-DEM simulations} \label{CoarseGrainingIBM}
\subsection{Theory}
Below, we present adaptations of the coarse-graining method of \S\ref{CoarseGraining} to the IBMs after \citet{Uhlmann05} and \citet{Bigotetal14}.

\subsubsection{IBM after Uhlmann (2005)} \label{TheoryIBM}
In the IBM after \citet{Uhlmann05}, the incompressible Navier-Stokes equations are solved in the entire fluid-particle domain $\mathbb{V}_\infty$, rather than only within $\mathbb{V}_\mathrm{f}$, replacing the particles' interiors by pseudo-fluid, while a singular force per unit area, $\bm{\tilde f}^p_\mathrm{IBM}[\bm{x}\in\mathbb{S}^p]$, enforces the (usually no-slip) boundary conditions at a particle $p$'s surface $\mathbb{S}^p$. This leads to a volumetric force density,
\begin{equation}
 \bm{f}_\mathrm{IBM}\equiv\sum\nolimits_p\int_{\mathbb{S}^p}\delta[\bm{x}-\bm{y}]\bm{\tilde f}^p_\mathrm{IBM}[\bm{y}]\d S_y,
\end{equation}
entering the microscopic fluid momentum balance, Eq.~(\ref{MomFluidmic}), as an additional body force term:
\begin{equation}
 \rho_\mathrm{f}(\partial_t\bm{\tilde u}+\bm{\tilde u}\cdot\bm{\nabla}\bm{\tilde u})=\bm{\nabla}\cdot\bm{\sigma}+\rho_\mathrm{f}\bm{g}+\bm{f}_\mathrm{IBM}. \label{MomIBM}
\end{equation}

In order to adapt the formalism developed in \S\ref{CoarseGraining} to Eq.~(\ref{MomIBM}), one first needs to circumvent the mathematical problem that the generalized-function products $X^p[\bm{x}]\delta[\bm{x}-\bm{y}]$ and $X_\mathrm{f}[\bm{x}]\delta[\bm{x}-\bm{y}]$ are not well-defined for $\bm{y}\in\mathbb{S}^p$. To this end, we introduce an infinitesimally extended particle $p$, occupying a domain $\mathbb{V}^p_+$ that is constructed such that the surface $\mathbb{S}^p$ of $\mathbb{V}^p$ lies strictly in the interior of $\mathbb{V}^p_+$. For example, for a spherical particle $p$ of radius $R_p$, integration over the volume $\mathbb{V}^p_+$ would then be equivalent to integration over a spherical domain of radius $(1+\epsilon)R_p$ and performing the limit $\epsilon\to0^+$ afterward. Likewise, we define the infinitesimally contracted domain $\mathbb{V}_{\mathrm{f}-}$ such that $\mathbb{S}^p$ lies strictly outside of $\mathbb{V}_{\mathrm{f}-}$. The corresponding generalized-function products therefore satisfy $X^p_+[\bm{x}]\delta[\bm{x}-\bm{y}]=\delta[\bm{x}-\bm{y}]$ and $X_{\mathrm{f}-}[\bm{x}]\delta[\bm{x}-\bm{y}]=0$, respectively, for $\bm{y}\in\mathbb{S}^p$. These relations imply that $\bm{f}_\mathrm{IBM}$ does not appear as an additional body force term in the macroscopic fluid phase momentum balance and that the hydrodynamic force acting on a particle $p$, using Eq.~(\ref{MomIBM}) and Reynolds transport theorem, can be calculated as\footnote{The equality in the second line of Eq.~(\ref{FhIBM}) assumes $\int_{\mathbb{V}^p_+}\sum_q\int_{\mathbb{S}^q}\delta[\bm{x}-\bm{y}]\bm{\tilde f}^q_\mathrm{IBM}[\bm{y}]\d S_y\d V_x=\int_{\mathbb{S}^p}\bm{\tilde f}^p_\mathrm{IBM}\d S$, which neglects the fact that particles are allowed to very slightly overlap. The precise calculation would give nonzero contributions not only for $q=p$ but also for $q\ne p$ if a portion of the surface $\mathbb{S}^q$ of a particle $q$ is within the extended domain $\mathbb{V}^p_+$ of a particle $p$.}
\begin{equation}
\begin{split}
 \bm{F}^p_\mathrm{h}&=\int_{\mathbb{S}^p_+}\bm{n}^p_+\cdot\bm{\sigma}\d S=\int_{\mathbb{V}^p_+}\bm{\nabla}\cdot\bm{\sigma}\d V \\
 &=\d_t\int_{\mathbb{V}^p}\rho_\mathrm{f}\bm{\tilde u}\d V-\int_{\mathbb{S}^p}\bm{\tilde f}^p_\mathrm{IBM}\d S-\rho_\mathrm{f}\bm{g}V^p,
\end{split} \label{FhIBM}
\end{equation}
where $\mathbb{S}^p_+=\partial\mathbb{V}^p_+$ is the infinitesimally extended surface domain of a particle $p$ and $\bm{n}^p_+$ its outward-directed unit normal vector. The bottom line of Eq.~(\ref{FhIBM}) is precisely how $\bm{F}^p_\mathrm{h}$ is calculated in Uhlmann's IBM~\citep{Uhlmann05,KempeFrohlich12b,Biegertetal17,Tschisgaleetal17,Zhuetal22}, considering that these studies modeled the surface $\mathbb{S}^p$ as a very thin volume layer, the so-called Lagrangian layer~\citep{Tschisgaleetal17}, extending minimally both into the interior and exterior of the particle $p$. This layer was then discretized into small grid cells, so that in total:
\begin{equation}
 \int_{\mathbb{S}^p}\bm{\tilde f}^p_\mathrm{IBM}\d S\simeq\sum\nolimits_l\bm{f}^p_l\Delta V^p_l, \label{LagrangianDiscretization}
\end{equation}
where $\bm{f}^p_l\Delta V^p_l$ is the force applied on the grid cell volume $\Delta V^p_l$ corresponding to a given Lagrange point (index $l$) on the surface of the particle $p$. In Eq.~(\ref{LagrangianDiscretization}), and also in other equations, such as those of Appendix~\ref{Implementation}, the symbol `$\simeq$' shall indicate approximations associated with DNS- or IBM-related discretizations.

The macroscopic fluid phase momentum balance resulting from Eqs.~(\ref{MomIBM}) and (\ref{FhIBM}) is still given by Eq.~(\ref{MomFluid}), but with a slightly modified expression for $\bm{\sigma^\mathrm{f}_\mathrm{s}}$, which is now based on the surface domain $\mathbb{S}^p_+$ rather than $\mathbb{S}^p$, and therefore for $\bm{\sigma^\mathrm{f}}$ in Eq.~(\ref{SigmaFluid}):
\begin{equation}
 \bm{\sigma^\mathrm{f}_\mathrm{s}}=\left\langle\sum\nolimits_p\left(S^p\underline{\bm{r}^p\bm{n}^p_+\cdot\bm{\sigma}\delta^p_{\mathrm{L}\bm{x}}}_{\mathbb{S}^p_+}-\underline{\bm{r}^p\delta^p_{\mathrm{L}\bm{x}}}_{\mathbb{V}^p}\bm{F}^p_\mathrm{h}\right)\right\rangle. \label{SigmaSIBM}
\end{equation}
In contrast, there is no change to the macroscopic solid phase momentum balance and the solid phase stresses, Eqs.~(\ref{MomSolid})-(\ref{SigmaSolidContact}).

One is now confronted with the difficulty that the stress tensor $\bm{\sigma}$, required to evaluate the right-hand side of Eq.~(\ref{SigmaSIBM}), is \textit{not} provided by the IBM, since the numerical algorithm does not solve Eq.~(\ref{MomIBM}). Instead, it solves a smoothed version of it based on a modified stress tensor $\bm{\tilde\sigma}$:
\begin{equation}
 \rho_\mathrm{f}(\partial_t\bm{\tilde u}+\bm{\tilde u}\cdot\bm{\nabla}\bm{\tilde u})=\bm{\nabla}\cdot\bm{\tilde\sigma}+\rho_\mathrm{f}\bm{g}+\overline{\bm{f}}_\mathrm{IBM}, \label{MomIBMsmooth}
\end{equation}
where $\overline{\bm{f}}_\mathrm{IBM}$ is a smoothed volumetric force density, defined as
\begin{equation}
\begin{split}
 \overline{\bm{f}}_\mathrm{IBM}&\equiv\int_{\mathbb{R}^3}\bm{f}_\mathrm{IBM}[\bm{x}-\bm{y}]\mathcal{G}_\Delta[\bm{y}]\d^3y \\
 &=\sum\nolimits_p\int_{\mathbb{S}^p}\mathcal{G}_\Delta[\bm{x}-\bm{y}]\bm{\tilde f}^p_\mathrm{IBM}[\bm{y}]\d S_y. \label{fIBMsmooth}
\end{split}
\end{equation}
Here $\mathcal{G}_\Delta[\bm{x}]$ is a smoothing function that satisfies $\int_{\mathbb{R}^3}\mathcal{G}_\Delta[\bm{x}]\d x^3=1$ and exhibits a compact support on the order of the DNS grid length $\Delta$ (Appendix~\ref{SmoothingFunction}). The key problem now is that $\mathcal{G}_\Delta$ spreads the particle surface force density $\bm{\tilde f}^p_\mathrm{IBM}$ onto the stationary Eulerian DNS grid, independently of $\Delta$ about equally to the interior and exterior of $\mathbb{V}^p_+$, with the result that $|\int_{\mathbb{V}^p_+}\overline{\bm{f}}_\mathrm{IBM}\d V|$ substantially underestimates $|\int_{\mathbb{V}^p_+}\bm{f}_\mathrm{IBM}\d V|$, leading to potentially substantial differences between $\int_{\mathbb{S}^p_+}\bm{n}^p_+\cdot\bm{\tilde\sigma}\d S$ and $\bm{F}^p_\mathrm{h}$. This issue remains present even in the limit $\Delta\to0^+$, implying that it cannot be dismissed as a discretization error. It suggests that Eq.~(\ref{SigmaSIBM}) may not be well approximated when the unknown $\bm{\sigma}$ on its right-hand side is approximated by $\bm{\tilde\sigma}$. In addition, in order to evaluate Eq.~(\ref{SigmaSIBM}), one would then need to interpolate the values of $\bm{\tilde\sigma}$ back from the DNS grid to the particle surfaces, creating another potential source of inaccuracy.

To solve both problems, we exploit the fact that the fluid fields are well-defined in the entire fluid-particle domain $\mathbb{V}_\infty$. Using the relations derived in Appendix~\ref{Identities} and $\bm{\nabla}\cdot\bm{\sigma}=\bm{\nabla}\cdot\bm{\tilde\sigma}+\overline{\bm{f}}_\mathrm{IBM}-\bm{f}_\mathrm{IBM}$, this permits calculating\footnote{The last equality in Eq.~(\ref{help1}) neglects the fact that particles are allowed to very slightly overlap. It is possible to account for particle overlap through replacing $S^p\underline{\bm{r}^p\bm{\tilde f}^p_\mathrm{IBM}\delta^p_{\mathrm{L}\bm{x}}}_{\mathbb{S}^p}$ by $X^p_+\sum\nolimits_qS^q\underline{\bm{r}^p\bm{\tilde f}^q_\mathrm{IBM}\delta^p_{\mathrm{L}\bm{x}}}_{\mathbb{S}^q}$ in Eq.~(\ref{help1}), and subsequently in Eq.~(\ref{SigmaIBMf}), which gives nonzero contributions not only for $q=p$ but also for $q\ne p$ if a portion of the surface $\mathbb{S}^q$ of a particle $q$ is within the extended domain $\mathbb{V}^p_+$ of a particle $p$. However, this would be inconsistent with the fact the we already neglected particle overlap in Eq.~(\ref{FhIBM}). We confirmed that, for our simulations presented in \S\ref{TestImplementation}, the effect of particle overlap on $\bm{\sigma^\mathrm{f}}$ is indeed negligible.}
\begin{equation}
\begin{split}
 &\bm{\nabla}\cdot S^p\underline{\bm{r}^p\bm{n}^p_+\cdot\bm{\sigma}\delta^p_{\mathrm{L}\bm{x}}}_{\mathbb{S}^p_+} \\
 =&\;\bm{\nabla}\cdot\left(X^p_+\bm{\sigma}+V^p\underline{\bm{r}^p(\bm{\nabla}\cdot\bm{\sigma})\delta^p_{\mathrm{L}\bm{x}}}_{\mathbb{V}^p_+}\right) \\
 =&\;\bm{\nabla}\cdot\left(X^p_+\bm{\sigma}+V^p\underline{\bm{r}^p\left(\bm{\nabla}\cdot\bm{\tilde\sigma}+\overline{\bm{f}}_\mathrm{IBM}\right)\delta^p_{\mathrm{L}\bm{x}}}_{\mathbb{V}^p}\right. \\
 &\quad\quad\quad\quad\;\;\,-\left.S^p\underline{\bm{r}^p\bm{\tilde f}^p_\mathrm{IBM}\delta^p_{\mathrm{L}\bm{x}}}_{\mathbb{S}^p}\right),
\end{split} \label{help1}
\end{equation}
Furthermore, since $\mathcal{G}_\Delta$ integrates to unity, the difference $\bm{\nabla}\cdot\bm{\sigma}-\bm{\nabla}\cdot\bm{\tilde\sigma}$ can be directly calculated from the surface force density $\bm{\tilde f}^p_\mathrm{IBM}$ as
\begin{equation}
\begin{split}
 &\bm{\nabla}\cdot\bm{\sigma}-\bm{\nabla}\cdot\bm{\tilde\sigma}=\overline{\bm{f}}_\mathrm{IBM}-\bm{f}_\mathrm{IBM} \\
 &=\int_{\mathbb{R}^3}\left(\bm{f}_\mathrm{IBM}[\bm{x}-\bm{r}]-\bm{f}_\mathrm{IBM}[\bm{x}]\right)\mathcal{G}_\Delta[\bm{r}]\d^3r \\
 &=\int_{\mathbb{R}^3}\int_0^1\frac{\d}{\d\lambda}\bm{f}_\mathrm{IBM}[\bm{x}-\lambda\bm{r}]\d\lambda\mathcal{G}_\Delta[\bm{r}]\d^3r \\
 &=-\bm{\nabla}\cdot\left(\int_{\mathbb{R}^3}\bm{r}\sum\nolimits_p\int_{\mathbb{S}^p}\bm{\tilde f}^p_\mathrm{IBM}[\bm{y}]\right. \\
 &\quad\quad\quad\left.\int_0^1\delta[\bm{x}-\bm{y}-\lambda\bm{r}]\d\lambda\d S_y\mathcal{G}_\Delta[\bm{r}]\d^3r\right).
\end{split} \label{help2}
\end{equation}
Inserting Eqs.~(\ref{help1}) and (\ref{help2}) into Eq.~(\ref{SigmaSIBM}) and the result into Eq.~(\ref{SigmaFluid}), using $\beta_\mathrm{f}\langle\bm{\sigma}\rangle^\mathrm{f}+\beta_\mathrm{s}\langle\bm{\sigma}\rangle^\mathrm{s}=\langle\bm{\sigma}\rangle$, we finally obtain IBM-adapted expressions for the effective fluid phase stress tensor,
\begin{align}
\begin{split}
 &\bm{\sigma^\mathrm{f}}=\langle\bm{\tilde\sigma}\rangle+\bm{\sigma^\mathrm{f}_\mathrm{Re}}+\bm{\sigma^\mathrm{f}_\mathrm{\tilde s}}+\bm{\sigma^\mathrm{f}_\Delta} \\
 &~~~\,\simeq\langle\bm{\tilde\sigma}\rangle+\bm{\sigma^\mathrm{f}_\mathrm{Re}}+\bm{\sigma^\mathrm{f}_\mathrm{\tilde s}},
\end{split}\label{SigmaIBM} \\
\begin{split}
 &\bm{\sigma^\mathrm{f}_\mathrm{\tilde s}}\equiv\left\langle\sum\nolimits_p\left(V^p\underline{\bm{r}^p\left(\bm{\nabla}\cdot\bm{\tilde\sigma}+\overline{\bm{f}}_\mathrm{IBM}\right)\delta^p_{\mathrm{L}\bm{x}}}_{\mathbb{V}^p}-\right.\right. \\
 &\left.\left.S^p\underline{\bm{r}^p\bm{\tilde f}^p_\mathrm{IBM}\delta^p_{\mathrm{L}\bm{x}}}_{\mathbb{S}^p}-\underline{\bm{r}^p\delta^p_{\mathrm{L}\bm{x}}}_{\mathbb{V}^p}\bm{F}^p_\mathrm{h}\right)\right\rangle,
\end{split} \label{SigmaIBMf} \\
\begin{split}
 &\bm{\sigma^\mathrm{f}_\Delta}\equiv\left\langle\int_{\mathbb{R}^3}\bm{r}\sum\nolimits_p\int_{\mathbb{S}^p}\bm{\tilde f}^p_\mathrm{IBM}[\bm{y}]\right. \\
 &\left.\int_0^1\delta[\bm{x}-\bm{y}-\lambda\bm{r}]\d\lambda\d S_y\mathcal{G}_\Delta[\bm{r}]\d^3r\right\rangle,
\end{split}
\end{align}
and the effective mixture stress tensor,
\begin{equation}
\begin{split}
 \bm{\sigma^\mathrm{m}}&=\langle\bm{\tilde\sigma}\rangle+\bm{\sigma^\mathrm{f}_\mathrm{\tilde s}}+\bm{\sigma^\mathrm{f}_\Delta}+\bm{\sigma^\mathrm{s}_\mathrm{c}}+\bm{\sigma^\mathrm{m}_\mathrm{Re}} \\
 &\simeq\langle\bm{\tilde\sigma}\rangle+\bm{\sigma^\mathrm{f}_\mathrm{\tilde s}}+\bm{\sigma^\mathrm{s}_\mathrm{c}}+\bm{\sigma^\mathrm{m}_\mathrm{Re}}.
\end{split} \label{SigmaIBMMixture}
\end{equation}
In Eqs.~(\ref{SigmaIBM}) and (\ref{SigmaIBMMixture}), we neglected the contribution $\bm{\sigma^\mathrm{f}_\Delta}$, since it vanishes in the limit $\Delta\to0^+$ due to $\mathcal{G}_\Delta[\bm{r}]\to\delta[\bm{r}]$, meaning that $\bm{\sigma^\mathrm{f}_\Delta}$ is a true artifact of the DNS- and IBM-related discretizations\footnote{We confirmed that, for our simulations presented in \S\ref{TestImplementation}, $\bm{\sigma^\mathrm{f}_\Delta}$ is indeed negligible.}. Hence, the crucial effect of the IBM-related smoothing with the function $\mathcal{G}_\Delta$ in Eqs.~(\ref{MomIBMsmooth}) and (\ref{fIBMsmooth}) is the intermingling of the Eulerian volumetric force density field $\overline{\bm{f}}_\mathrm{IBM}$ and the singular surface force $\bm{\tilde f}^p_\mathrm{IBM}$ per unit area in Eq.~(\ref{SigmaIBMf}), since this effect does not vanish in the limit $\Delta\to0^+$.

\subsubsection{IBM after Bigot et al. (2014)}
In the IBM after \citet{Bigotetal14}, the volumetric force density $\bm{f}_\mathrm{IBM}$ in Eq.~(\ref{MomIBM}) is not singular at the surface but continuous within each particle $p$'s volume $\mathbb{V}^p$. It is constructed to not only enforce no-slip boundary conditions at its surface $\mathbb{S}^p$ but to ensure equality between the pseudo-fluid velocity $\bm{\tilde u}$ and the local particle velocity $\bm{v}^p_\uparrow+\bm{\omega}^p\times\bm{r}^p$ in its entire domain $\mathbb{V}^p$. Hence, one could, in principle, use Eqs.~(\ref{SigmaFluid}) and (\ref{SigmaMixture}) to calculate the effective stress tensors $\bm{\sigma^\mathrm{f}}$ and $\bm{\sigma^\mathrm{m}}$. However, based on an equivalent analysis to that in \S\ref{TheoryIBM}, more accurate expressions that avoid interpolating $\bm{\sigma}$ to the particle surfaces are given by
\begin{align}
 \bm{\sigma^\mathrm{f}}=&\;\langle\bm{\sigma}\rangle+\bm{\sigma^\mathrm{f}_\mathrm{Re}}+\bm{\sigma^\mathrm{f}_\mathrm{\hat s}}, \\
\begin{split}
 \bm{\sigma^\mathrm{f}_\mathrm{\hat s}}\equiv&\;\left\langle\sum\nolimits_p\left(V^p\underline{\bm{r}^p(\bm{\nabla}\cdot\bm{\sigma})\delta^p_{\mathrm{L}\bm{x}}}_{\mathbb{V}^p}\right.\right. \\
 &-\left.\left.\underline{\bm{r}^p\delta^p_{\mathrm{L}\bm{x}}}_{\mathbb{V}^p}\bm{F}^p_\mathrm{h}\right)\right\rangle,
\end{split} \\
 \bm{\sigma^\mathrm{m}}=&\;\langle\bm{\sigma}\rangle+\bm{\sigma^\mathrm{f}_\mathrm{\hat s}}+\bm{\sigma^\mathrm{s}_\mathrm{c}}+\bm{\sigma^\mathrm{m}_\mathrm{Re}}.
\end{align}

\subsection{Test with simulations of fluid-driven sediment transport} \label{TestImplementation}
\begin{figure}
 \includegraphics[width=\columnwidth]{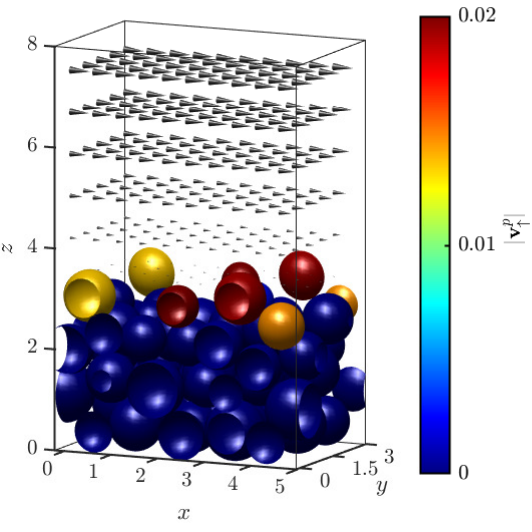}
\caption{Snapshot of a DNS-DEM simulation of steady, bed-tangentially homogeneous sediment transport.}
\label{Sketch}
\end{figure}

We carry out IBM-based DNS-DEM simulations of steady, bed-tangentially homogeneous sediment transport driven by an incompressible Newtonian fluid (Fig.~\ref{Sketch}), a particulate two-phase flow with $R/L$ on the order of $1$ to $10$~\citep{Chassagneetal23,Tholenetal23,Fryetal24}.
The gravitational acceleration $\bm{g}$ is driving both the liquid flow and particles down a small slope of $3^\circ$. Physical quantities are measured in natural units of $\rho_\mathrm{f}$, $(\rho_\mathrm{s}/\rho_\mathrm{f}-1)|\bm{g}|$, and $2R$. A simulation is conducted for $\rho_\mathrm{s}=2.65$ and a high viscosity $\eta_\mathrm{f}=0.36$, yielding a weakly inertial flow regime with a bulk flow Reynolds number $\mathrm{Re}=\rho_\mathrm{f}\langle u_\mathrm{max}\rangle_tH/\eta_\mathrm{f}\approx22.5$, where $\langle u_\mathrm{max}\rangle_t$ is the time-averaged maximum fluid velocity and $H$ is the extension of the vertical domain. The flow direction is $x$, the direction normal to the bed is $z$, and the lateral direction is $y$. Since we are primarily concerned with the validation of the derived micromechanical expressions, we keep the system size small, $(L_x,L_y,H)=(5,3,8)$, and consider only $50$ identical, spherical particles, forming a bed of approximately three particle layers (Fig.~\ref{Sketch}). Here $L_x$ and $L_y$ denote the length and width, respectively, of the $(x,y)$-domain periodic in both $x$ and $y$, while no-slip and free-slip boundary conditions are imposed at the bottom ($z=0$) and top ($z=H$) walls of the vertical domain to mimic a rigid and free surface, respectively. Note that much-larger-scale simulations of high-inertia flow are currently under way and will be presented in future studies.

The numerical model is described in detail by \citet{Biegertetal17} or \citet{Zhuetal22}, based on the IBM after \citet{Uhlmann05} (\S\ref{TheoryIBM}). The microscopic fluid momentum balance equation, Eq.~(\ref{MomIBMsmooth}), is integrated using a third-order low-storage Runge-Kutta scheme for time advancement and a finite-difference approach for spatial discretization. The pressure field is solved using a direct solver based on the Fast Fourier Transform. To integrate the equations governing the particles' motion, the Runge-Kutta scheme subdivides the three-step fluid integration procedure into a total of $15$ substeps per fluid time step. Hydrodynamic forces acting on particles are calculated using Eqs.~(\ref{FhIBM}) and (\ref{LagrangianDiscretization}), while contact forces are calculated using the soft-sphere model proposed by \citet{Biegertetal17}. When the distance between the surfaces of two approaching particles becomes sufficiently small, the intervening fluid is squeezed out of the gap. Once the gap width falls below approximately $2\Delta$, this process can no longer be resolved by the fluid grid. Therefore, a lubrication model, Eq.~(2.10) of \citet{Zhuetal22}, is employed to account for the unresolved hydrodynamic interactions. This model is also applied to particles rebounding after collision, when fluid is drawn back into the gap. The lubrication force is dissipative, as it always acts in the direction opposite to the relative velocity. Besides, the Adaptive Collision Time Model proposed by \citet{KempeFrohlich12a} is employed to account for normal contact interactions. It stretches the collision process in time to match the time step of the fluid solver.

The numerical implementation of the micromechanical expressions associated with the two-fluid mass and momentum balances derived in \S\ref{CoarseGraining}, using the solid phase stress tensor expressions in \S\ref{DerivationMomentum} and the IBM-adapted fluid phase and mixture stress tensor expressions in \S\ref{TheoryIBM}, is straightforward. In Appendix~\ref{Implementation}, all terms constituting these expressions are transformed into readily implementable forms using the following averaging procedure:
\begin{equation}
\begin{split}
 &\langle A\rangle[z_n]\equiv\frac{1}{N_TL_xL_y\Delta z}\times \\
 &\sum_{k=1}^{N_T}\int_{z_n-\Delta z/2}^{z_n+\Delta z/2}\int_0^{L_y}\int_0^{L_x}A[\bm{x},t_k]\d x\d y\d z, \label{AveragingDNSDEM}
\end{split}
\end{equation}
which satisfies the rules in Eq.~(\ref{AveragingProcedure}). Due to statistical bed-tangential homogeneity ($\partial_x=\partial_y=0$), the averaging takes place over the entire periodic bed-tangential domain, $x\in[0,L_x)$ and $y\in[0,L_y)$, and, due to statistical steadiness, over a sufficient number $N_T=268$ of timewise \textit{well-separated} instants $t_k$ to ensure $\partial_t\approx0$. Furthermore, $\Delta z=0.1$ is the constant extent of the intervals with centers $z_n$ that discretize the macroscopic bed-normal domain $z\in[0,H]$. It is twice the DNS grid length, $\Delta=0.05$, and sufficiently small to ensure that Eq.~(\ref{AveragingDNSDEM}) approximately satisfies the Reynolds rule, Eq.~(\ref{ReynoldsRule}).

Due to the regularity of the simulation domain and the spherical shape of the particles, the integrals in Eq.~(\ref{AveragingDNSDEM}) can be calculated analytically when weighted with the indicator function $X^p$ of a particle $p$, implying that coarse-grained particle-related fields can be obtained from the simulation data without discretization errors (Appendix~\ref{ParticleRelated}). In contrast, such errors cannot be avoided when calculating coarse-grained fluid-related quantities due to the involved DNS and IBM mesh grid spacing (Appendix~\ref{FluidRelated}).

\begin{figure*}
 \includegraphics[width=\textwidth]{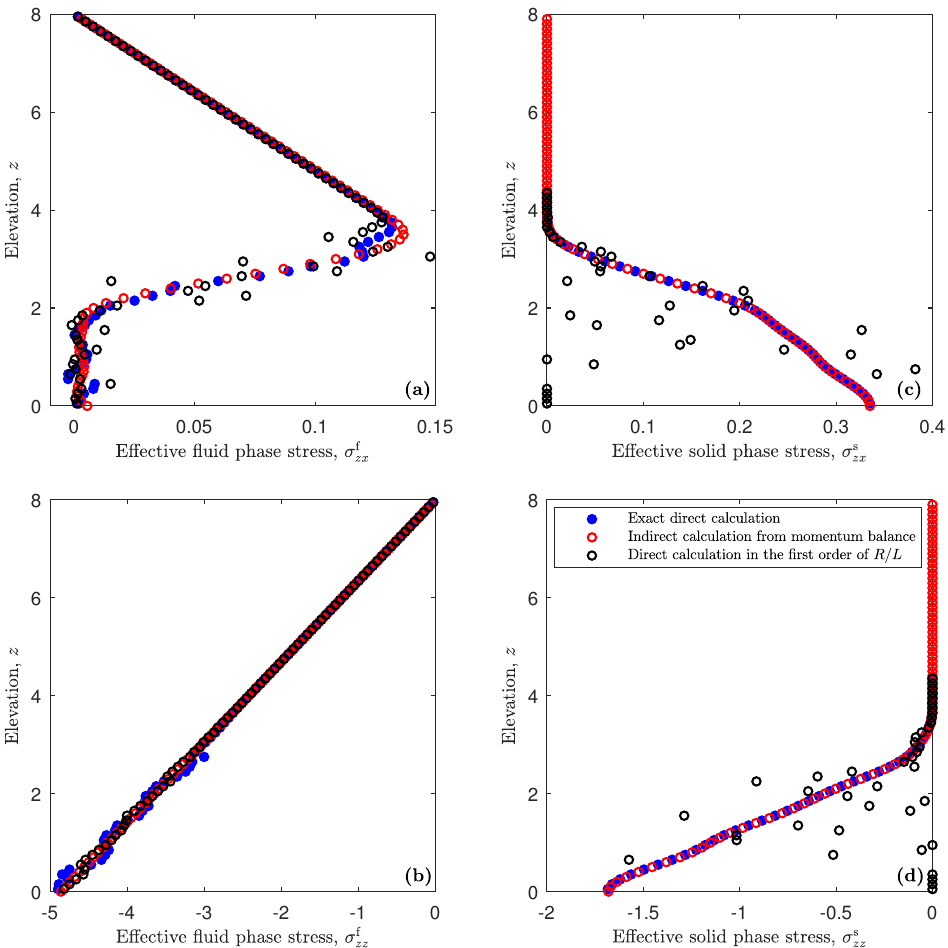}
\caption{Vertical profiles of the effective stress components (a) $\sigma^\mathrm{f}_{zx}$, (b) $\sigma^\mathrm{f}_{zz}$, (c) $\sigma^\mathrm{s}_{zx}$, and (d) $\sigma^\mathrm{s}_{zz}$. Data correspond to a IBM-based DNS-DEM simulation of sediment transport, calculated in three distinct manners: directly using the derived micromechanical expressions, Eqs.~(\ref{SigmaSolid}) and (\ref{SigmaIBM}), indirectly via integrating the right-hand sides of Eqs.~(\ref{SigmaFluidSteady}) and (\ref{SigmaSolidSteady}), or directly using the stress expressions approximated in the first order of $R/L$ (see Appendix~\ref{Implementation} for implementation details).}
\label{Stresses}
\end{figure*}

Because of $\partial_x=\partial_y=\partial_t=0$, the coupled macroscopic fluid and solid phase momentum balances take the simple forms
\begin{align}
 \d_z\sigma^\mathrm{f}_{zi}&=-\beta_\mathrm{f}\rho_\mathrm{f}g_i+\beta_\mathrm{s}\langle f_{\mathrm{h}i}\rangle^\mathrm{s}, \label{SigmaFluidSteady} \\
 \d_z\sigma^\mathrm{s}_{zi}&=-\beta_\mathrm{s}\rho_\mathrm{s}g_i-\beta_\mathrm{s}\langle f_{\mathrm{h}i}\rangle^\mathrm{s}. \label{SigmaSolidSteady}
\end{align}
In Fig.~\ref{Stresses}, we show the vertical profiles of the effective stress components $\sigma^\mathrm{f}_{zx}$, $\sigma^\mathrm{f}_{zz}$, $\sigma^\mathrm{s}_{zx}$, and $\sigma^\mathrm{s}_{zz}$ calculated in three distinct manners: directly using the derived micromechanical stress expressions, Eqs.~(\ref{SigmaSolid}) and (\ref{SigmaIBM}), indirectly via integrating the right-hand sides of Eqs.~(\ref{SigmaFluidSteady}) and (\ref{SigmaSolidSteady}), or directly using the stress expressions approximated in the first order of $R/L$ (see Appendix~\ref{Implementation} for implementation details).
The latter approximation (Appendix~\ref{ComparisonFry}) is carried out to obtain a coarse-graining formulation equivalent to that of \citet{Fryetal24}, who also simulated steady, bed-tangentially homogeneous sediment transport in a weakly inertial regime, but using a Couette flow geometry. It can be seen that the exact direct and indirect calculations are approximately equal to each another, validating the derivations in \S\ref{DerivationMomentum} and \S\ref{TheoryIBM}, respectively. However, the first-order-approximated formulation fails for $\sigma^\mathrm{f}_{zx}$, $\sigma^\mathrm{s}_{zx}$, and $\sigma^\mathrm{s}_{zz}$, as expected due to the lack of scale separation between $R$ and $L$. In contrast to us, \citet{Fryetal24} reported good agreement between their indirect and approximated direct expressions. However, even though they conducted simulations for a large range of flow-driving shear forces, their coarse-graining formulation was validated only for their most intense flow condition. The intenser the flow, the larger is $L$ and the better becomes the first-order approximation in $R/L$. In our simulations, the flow driving is comparably much weaker. Furthermore, \citet{Fryetal24} used an averaging procedure different from Eq.~(\ref{AveragingDNSDEM}). While they also averaged over $x$, $y$, and $t$, their vertical averaging was based on a Gaussian kernel with a standard deviation $\Delta z=1/\sqrt{2}$. Such a large coarse-graining scale smooths out much of the variability of the \textit{macroscopic} flow, such as the relatively very strong change of $\sigma^\mathrm{f}_{zx}$ occurring between about $z=2$ and $z=3$ in Fig.~\ref{Stresses}(a). In fact, since every elevation $z$ in any given steady, homogeneous inclined-flow configuration (not just sediment transport) corresponds to a different macroscopic flow state, one ideally wants $\Delta z$ to be as small as possible.

\section{Conclusions}
We have presented a general method to coarse-grain the equations of motion of a mixture of a continuous fluid of arbitrary rheology and non-Brownian particles, interacting via contacts, of arbitrary shape and composition (Supplementary Material), and exemplary applied it to derive the coupled two-fluid mass and momentum balances of a mixture of an incompressible fluid and particles of constant density (\S\ref{CoarseGraining}). The method's key novelties are as follows: First, the derived micromechanical expressions for the macroscopic fields of interest, such as the effective fluid and solid phase stresses in Eqs.~(\ref{SigmaFluid}) and (\ref{SigmaSolid}), respectively, are mathematically exact for any ratio between the typical particle size $R$, or the typical particle separation distance, and the macroscopic flow scale $L$, and can be readily extracted from DNS-DEM simulations in a computationally cheap manner (e.g., Appendix~\ref{Implementation}). In contrast, previous exact expressions contain infinite sums of derivatives of ever higher order~\citep{Jackson97,ZhangProsperetti97,FintziPierson26}. Their extraction from DNS-DEM data is not only inconvenient but also computationally costly, since ever finer grids are required to reliably calculate these increasing-order derivatives. Second, the microscopic volume fraction of each particle is its corresponding indicator function, rather than the traditional volume-weighted delta distribution at its center of mass, to ensure that the resulting macroscopic fluid and solid volume fractions add precisely to unity. This leads to an additional contact stress contribution not seen in standard coarse-grained expressions for granular matter (see Eq.~(\ref{SigmaSolidContact})), and to particle-rotational contributions to translational macroscopic solid phase balance equations (e.g., see Eq.~(\ref{SigmaSolidReynolds})). As a side note, the fact that these particle-rotational contributions vanish identically for spherical particles suggests that the coarse-graining method may potentially also find application when studying grain shape effects in pure granular flows.

Furthermore, we have presented mathematically exact adaptations of the method that are applicable to the IBM after \citet{Uhlmann05} and the IBM after \citet{Bigotetal14}, respectively, and implemented the former to obtain coarse-grained fields from DNS-DEM simulations of steady, homogeneous sediment transport based on the same IBM (\S\ref{CoarseGrainingIBM}). Sediment transport represents a particulate two-phase flow with $R/L$ on the order of $1$ to $10$~\citep{Chassagneetal23,Tholenetal23,Fryetal24}. To validate the implementation, we have shown that the extracted effective fluid and solid phase stresses are about equal to those obtained indirectly from integrating the respective momentum balances. In contrast, expressions approximated in the first order of $R/L$, like those used in a recent previous study~\citep{Fryetal24}, are in strong disagreement with the exact indirectly calculated ones.

In the future, we intend to use the coarse-graining method and its IBM adaptation to study the rheology of particulate two phase flows that lack scale separation using IBM-based DNS-DEM simulations.

\backmatter

\bmhead{Supplementary information}
The online version contains supplementary material available at \url{https://doi.org/10.1007/s10035-025-01609-5}.

\bmhead{Code availability}
A MATLAB code calculating the coarse-grained fields from the raw data produced by the IBM-based DNS–DEM simulations of sediment transport is available at \url{https://doi.org/10.5281/zenodo.18146169}.

\bmhead{Acknowledgments}
Z.H. acknowledges financial support from grant National Key R \& D Program of China (2023YFC3008100). T.P. acknowledges financial support from grants National Natural Science Foundation of China (nos.~12350710176, 12272344). R.Z. acknowledges financial support from grant National Natural Science Foundation of China (no.~12402494). Z.H. acknowledges financial support from grant National Natural Science Foundation of China (no.~52171276).

\bmhead{Author contributions}
Conceptualization: T.P., K.T.; Formal analysis: T.P., R.Z., Y.C.; Funding acquisition: T.P., Z.H., R.Z.; Investigation: T.P., R.Z., Y.C.; Methodology: T.P., R.Z., K.T.; Project administration: T.P., Z.H.; Resources: T.P., Z.H., R.Z.; Software: R.Z., T.P.; Supervision: T.P., Z.H.; Validation: T.P., R.Z., Y.C.; Visualization: T.P., R.Z., Y.C.; Writing - original draft: T.P.; Writing - review \& editing: T.P., Y.C, R.Z., K.T:

\bmhead{Data availability}
The data shown in Fig. 3 can be produced using the MATLAB source files uploaded at \url{https://doi.org/10.5281/zenodo.18146169}.

\section*{Declarations}
\bmhead{Conflict of interest}
The authors declare that they have no conflict of interest.

\bmhead{Ethics approval}
Not applicable.

\begin{appendices}

\section{Mathematical identities} \label{Identities}
Like previous studies~\citep{Babic97,Goldhirsch10}, we exploit the identity
\begin{equation}
 \delta[\bm{x}-\bm{y}]=\delta[\bm{x}-\bm{x}^p]-\bm{\nabla_x}\cdot(\bm{r}^p\delta^p_{\mathrm{L}\bm{x}})[\bm{y}], \label{DeltaId}
\end{equation}
with $\delta^p_{\mathrm{L}\bm{x}}[\bm{y}]$ given by Eq.~(\ref{DeltaLine}). Using it, we calculate:
\begin{equation}
\begin{split}
 &AX^p=\int_{\mathbb{R}^3}(AX^p)[\bm{y}]\delta[\bm{x}-\bm{y}]\d^3y \\
 &=\delta[\bm{x}-\bm{x}^p]\int_{\mathbb{V}^p}A\d V-\bm{\nabla}\cdot\int_{\mathbb{V}^p}\bm{r}^pA\delta^p_{\mathrm{L}\bm{x}}\d V \\
 &=\underline{A}_{\mathbb{V}^p}V^p\delta[\bm{x}-\bm{x}^p]-\bm{\nabla}\cdot V^p\underline{\bm{r}^pA\delta^p_{\mathrm{L}\bm{x}}}_{\mathbb{V}^p}, \label{IdentityV}
\end{split}
\end{equation}
where $A$ is a field well-defined within $\mathbb{V}^p$. In particular, the case $A=1$ implies
\begin{equation}
 V^p\delta[\bm{x}-\bm{x}^p]=X^p+\bm{\nabla}\cdot V^p\underline{\bm{r}^p\delta^p_{\mathrm{L}\bm{x}}}_{\mathbb{V}^p}, \label{XpDelta}
\end{equation}
from which we conclude
\begin{equation}
 AX^p=\underline{A}_{\mathbb{V}^p}X^p-\bm{\nabla}\cdot V^p\underline{\bm{r}^p\left(A-\underline{A}_{\mathbb{V}^p}\right)\delta^p_{\mathrm{L}\bm{x}}}_{\mathbb{V}^p}.
\end{equation}
Analogously, using $\bm{\nabla}X^p[\bm{x}]=-\bm{n}^p[\bm{x}]\int_{\mathbb{S}^p}\delta[\bm{x}-\bm{y}]\d S_y$~\citep{Drew83}, we calculate:
\begin{equation}
\begin{split}
 &A\bm{\nabla}X^p=\int_{\mathbb{R}^3}(A\bm{\nabla}X^p)[\bm{y}]\delta[\bm{x}-\bm{y}]\d^3y \\
 =&-\delta[\bm{x}-\bm{x}^p]\int_{\mathbb{S}^p}A\bm{n}^p\d S+\bm{\nabla}\cdot\int_{\mathbb{S}^p}\bm{r}^pA\bm{n}^p\delta^p_{\mathrm{L}\bm{x}}\d S \\
 =&-\underline{A\bm{n}^p}_{\mathbb{S}^p}S^p\delta[\bm{x}-\bm{x}^p]+\bm{\nabla}\cdot S^p\underline{\bm{r}^pA\bm{n}^p\delta^p_{\mathrm{L}\bm{x}}}_{\mathbb{S}^p} \\
 =&-\frac{S^p}{V^p}\underline{A\bm{n}^p}_{\mathbb{S}^p}X^p\\
 &+\bm{\nabla}\cdot S^p\left(\underline{\bm{r}^pA\bm{n}^p\delta^p_{\mathrm{L}\bm{x}}}_{\mathbb{S}^p}-\underline{\bm{r}^p\delta^p_{\mathrm{L}\bm{x}}}_{\mathbb{V}^p}\underline{A\bm{n}^p}_{\mathbb{S}^p}\right), \label{IdentityS}
\end{split}
\end{equation}
from which follow Eqs.~(\ref{InteractionTerm}) and (\ref{InteractionTerm2}).

\section{IBM smoothing function} \label{SmoothingFunction}
The smoothing function $\mathcal{G}_\Delta[\bm{x}]$ is defined as
\begin{equation}
 \mathcal{G}_\Delta[\bm{x}]\equiv\frac{1}{\Delta^3}\mathcal{G}[x/\Delta]\mathcal{G}[y/\Delta]\mathcal{G}[z/\Delta],
\end{equation}
with
\begin{equation}
 \mathcal{G}[\lambda]\equiv
 \begin{cases}
  \frac{1}{3}\left(1+\sqrt{1-3\lambda^2}\right)&\!\!\!\!\!\!\!\!|\lambda|\leq\frac{1}{2} \\
	0&\!\!\!\!\!\!\!\!|\lambda|\geq\frac{3}{2} \\
	\frac{1}{6}\left(5-3|\lambda|-\sqrt{1-3(1-|\lambda|)^2}\right)&\text{else}.
 \end{cases}
\end{equation}

\section{Implementation of IBM-adapted coarse-graining method} \label{Implementation}
\subsection{Averaging notation}
To ease the notation, we will only perform spatial averaging,
\begin{equation}
\begin{split}
 &\langle A\rangle_t[z_n]\equiv\frac{1}{L_xL_y\Delta z}\times \\
 &\int_{z_n-\Delta z/2}^{z_n+\Delta z/2}\int_0^{L_y}\int_0^{L_x}A[\bm{x},t]\d x\d y\d z,
\end{split} \label{Average_xyz}
\end{equation}
since the additional time averaging is trivial:
\begin{equation}
 \langle A\rangle[z_n]=\frac{1}{N_T}\sum_{k=1}^{N_T}\langle A\rangle_{t_k}[z_n].
\end{equation}

\subsection{Analytical expressions for certain generalized functions}
Since the particles in the simulations are spheres, the indicator function of a particle $p$ with radius $R_p$ can be written as
\begin{equation}
 X^p=H[R_p-|\bm{r}^p|], \label{Heaviside}
\end{equation}
where $H$ is the Heaviside function. Using this expression and $V^p=4\pi R_p^3/3$, one can calculate
\begin{equation}
\begin{split}
 &\frac{4\pi R_p^3}{3}\underline{\bm{r}^p\delta^p_{\mathrm{L}\bm{x}}}_{\mathbb{V}^p} \\
 &=\int_{\mathbb{V}^p}\bm{r}^p[\bm{y}]\int_0^1\delta[\bm{x}-\bm{x}^p-\lambda(\bm{y}-\bm{x}^p)]\d\lambda\d V_y \\
 &=\int_0^1\int_{\mathbb{R}^3}\bm{r}H[R_p-|\bm{r}|]\delta[\bm{r}^p-\lambda\bm{r}]\d^3r\d\lambda \\
 &=\bm{r}^p\int_0^1\lambda^{-4}H[R_p-|\bm{r}^p|/\lambda]\d\lambda \\
 &=H[R_p-|\bm{r}^p|]\bm{r}^p\int_{|\bm{r}^p|/R_p}^1\lambda^{-4}\d\lambda \\
 &=\frac{H[R_p-|\bm{r}^p|]\bm{r}^p}{3R_p}\left(\left(\frac{R_p}{|\bm{r}^p|}\right)^3-1\right).
\end{split} \label{rp_delta}
\end{equation}

\subsection{Coarse-grained particle-related quantities} \label{ParticleRelated}
Introducing polar coordinates, $\bm{r}^p=r\cos\phi\bm{e_x}+r\sin\phi\bm{e_y}+(z-z^p)\bm{e_z}$, with $\bm{e_x}$, $\bm{e_y}$, and $\bm{e_z}$ the unit vectors of the Cartesian coordinate system, using Eqs.~(\ref{Heaviside}) and (\ref{rp_delta}), it is now straightforward to carry out the integration in Eq.~(\ref{Average_xyz}) and therefore transform the coarse-grained particle-related quantities required in \S\ref{TestImplementation} into a readily implementable form:
\begin{align}
\begin{split}
 &\langle X_\mathrm{s}\rangle_t[z_n]=\frac{\pi}{V_z}\times \\
 &\sum_{p:z_{n2}^p>z_{n1}^p}\left[R_p^2z-\frac{1}{3}(z-z^p)^3\right]_{z=z_{n1}^p}^{z=z_{n2}^p},
\end{split} \\
\begin{split}
 &\langle X_\mathrm{s}\langle\bm{u}_\uparrow\rangle^\mathrm{s}\rangle_t[z_n]=\frac{\pi}{V_z}\times \\
 &\sum_{p:z_{n2}^p>z_{n1}^p}\bm{v}_\uparrow^p\left[R_p^2z-\frac{1}{3}(z-z^p)^3\right]_{z=z_{n1}^p}^{z=z_{n2}^p},
\end{split} \\
\begin{split}
 &\langle X_\mathrm{s}\langle\bm{u}_\uparrow\bm{u}_\uparrow\rangle^\mathrm{s}\rangle_t[z_n]=\frac{\pi}{V_z}\times \\
 &\sum_{p:z_{n2}^p>z_{n1}^p}\bm{v}_\uparrow^p\bm{v}_\uparrow^p\left[R_p^2z-\frac{1}{3}(z-z^p)^3\right]_{z=z_{n1}^p}^{z=z_{n2}^p},
\end{split} \\
\begin{split}
 &\langle X_\mathrm{s}\bm{f}_\mathrm{h}\rangle_t[z_n]=\frac{\pi}{V_z}\times \\
 &\sum_{p:z_{n2}^p>z_{n1}^p}\bm{f}^p_\mathrm{h}\left[R_p^2z-\frac{1}{3}(z-z^p)^3\right]_{z=z_{n1}^p}^{z=z_{n2}^p},
\end{split} \\
\begin{split}
 &\left\langle\sum\nolimits_p\underline{\bm{r}^p\delta^p_{\mathrm{L}\bm{x}}}_{\mathbb{V}^p}\bm{F}^p_\mathrm{h}\right\rangle_t[z_n]=\frac{\bm{e_z}}{2R_p^3V_z}\otimes \\
 &\sum_{p:z_{n2}^p>z_{n1}^p}\bm{F}^p_\mathrm{h}\left[R_p^3|z-z^p|-\frac{3}{4}R_p^2(z-z^p)^2\right. \\
 &\quad\quad\quad\quad\quad\quad+\left.\frac{1}{8}(z-z^p)^4\right]_{z=z_{n1}^p}^{z=z_{n2}^p},
\end{split} \label{FhDelta} \\
\begin{split}
 &\left\langle\sum\nolimits_p\underline{\bm{r}^p\delta^p_{\mathrm{L}\bm{x}}}_{\mathbb{V}^p}\bm{F}^p_\mathrm{c}\right\rangle_t[z_n]=\frac{\bm{e_z}}{2R_p^3V_z}\otimes \\
 &\sum_{p:z_{n2}^p>z_{n1}^p}\bm{F}^p_\mathrm{c}\left[R_p^3|z-z^p|-\frac{3}{4}R_p^2(z-z^p)^2\right. \\
 &\quad\quad\quad\quad\quad\quad+\left.\frac{1}{8}(z-z^p)^4\right]_{z=z_{n1}^p}^{z=z_{n2}^p}, \label{FcDelta}
\end{split}
\end{align}
with
\begin{align}
 V_z&\equiv L_xL_y\Delta z, \\
 z_{n2}^p&\equiv\min(z_n+\Delta z/2,z^p+R_p), \\
 z_{n1}^p&\equiv\max(z_n-\Delta z/2,z^p-R_p).
\end{align}
Furthermore,
\begin{equation}
\begin{split}
 &\left\langle\frac{1}{2}\sum\nolimits_{pq}\left(\bm{r}^{pq}_\mathrm{c}\delta^{pq}_{\mathrm{cL}\bm{x}}-\bm{r}^{qp}_\mathrm{c}\delta^{qp}_{\mathrm{cL}\bm{x}}\right)\bm{F}^{pq}_\mathrm{c}\right\rangle_t[z_n] \\
 &=\frac{1}{2V_z}\sum\nolimits_{pq}(\bm{x}^q-\bm{x}^p)\bm{F}^{pq}_\mathrm{c}w^p_n[z^q],
\end{split} \label{SigmaC}
\end{equation}
where $w^p_n[z]$ is the fraction of the line connecting $z$ with $z^p$ that is contained within the discretization interval $[z_n-\Delta z/2,z_n+\Delta z/2)$:
\begin{equation}
w^p_n[z]=
 \begin{cases}
  0&\text{if}\quad l^p_n[z]<0 \\
	1&\text{if}\quad l^p_n[z]=0\land z=z^p \\
	\frac{l^p_n[z]}{|z-z^p|}&\text{else},
 \end{cases}
\end{equation}
with
\begin{equation}
\begin{split}
 l^p_n[z]\equiv&\;\min\left(\max(z,z^p),z_n+\Delta z/2\right) \\
 &-\max\left(\min(z,z^p),z_n-\Delta z/2\right).
\end{split}
\end{equation}

\subsection{Coarse-grained fluid-related quantities} \label{FluidRelated}
The coarse-grained fluid-related quantities required in \S\ref{TestImplementation} are numerically calculated via iterating through the DNS grid:
\begin{align}
 \langle X_\mathrm{f}\rangle_t[z_n]&\simeq\frac{1}{V_z}\sum_{\substack{z_\alpha\in[z_n-\Delta z/2,\\z_n+\Delta z/2)}}V_{\mathrm{f}\alpha}, \\
 \left\langle X_\mathrm{f}\bm{u}_\uparrow\right\rangle_t[z_n]&\simeq\frac{1}{V_z}\sum_{\substack{z_\alpha\in[z_n-\Delta z/2,\\z_n+\Delta z/2)}}V_{\mathrm{f}\alpha}\bm{\tilde u}_\alpha, \\
 \left\langle X_\mathrm{f}\bm{u}_\uparrow\bm{u}_\uparrow\right\rangle_t[z_n]&\simeq\frac{1}{V_z}\sum_{\substack{z_\alpha\in[z_n-\Delta z/2,\\z_n+\Delta z/2)}}V_{\mathrm{f}\alpha}\bm{\tilde u}_\alpha\bm{\tilde u}_\alpha, \\
 \langle\tilde\sigma\rangle_t[z_n]&\simeq\frac{1}{V_z}\sum_{\substack{z_\alpha\in[z_n-\Delta z/2,\\z_n+\Delta z/2)}}\Delta^3\bm{\tilde\sigma}_\alpha,
\end{align}
where the index $\alpha$ enumerates the grid cells of the DNS grid, and $V_{\mathrm{f}\alpha}$ and $\Delta^3$ are the fluid volume and total volume, respectively, contained within the grid cell $\alpha$. Furthermore,
\begin{align}
\begin{split}
 &\left\langle\sum\nolimits_pV^p\underline{\bm{r}^p\left(\bm{\nabla}\cdot\bm{\tilde\sigma}+\overline{\bm{f}}_\mathrm{IBM}\right)\delta^p_{\mathrm{L}\bm{x}}}_{\mathcal{V}^p}\right\rangle_t[z_n] \\
 &\simeq\frac{1}{V_z}\sum\nolimits_p\sum\nolimits_\alpha\bm{r}^p_\alpha\left(\bm{\nabla}\cdot\bm{\tilde\sigma}+\overline{\bm{f}}_\mathrm{IBM}\right)_\alpha w^p_n[z_\alpha]V^p_\alpha,
\end{split} \label{SigmaF1} \\
\begin{split}
 &\left\langle\sum\nolimits_pS^p\underline{\bm{r}^p\bm{\tilde f}^p_\mathrm{IBM}\delta^p_{\mathrm{L}\bm{x}}}_{\mathcal{S}^p}\right\rangle_t[z_n] \\
 &\simeq\frac{1}{V_z}\sum\nolimits_p\sum\nolimits_l\bm{r}^p_l\bm{\tilde f}^p_{\mathrm{IBM}l}w^p_n[z_l]S^p_l,
\end{split} \label{SigmaF20} \\
\begin{split}
 &\left\langle\sum\nolimits_pX^p_+\sum\nolimits_qS^q\underline{\bm{r}^p\bm{\tilde f}^q_\mathrm{IBM}\delta^p_{\mathrm{L}\bm{x}}}_{\mathcal{S}^q}\right\rangle_t[z_n] \\
 &\simeq\frac{1}{V_z}\sum_{pq:|\bm{r}^p_{l_q}|\leq R_p}\sum_{l_q}\bm{r}^p_{l_q}\bm{f}^q_{l_q}w^p_n[z_{l_q}]\Delta V^q_{l_q},
\end{split} \label{SigmaF2} \\
\begin{split}
 &\left\langle\int_{\mathcal{R}^3}\bm{r}\sum\nolimits_p\int_{\mathcal{S}^p}\bm{\tilde f}^p_\mathrm{IBM}[\bm{y}]\right. \\
 &\left.\int_0^1\delta[\bm{x}-\bm{y}-\lambda\bm{r}]\d\lambda\d S_y\mathcal{G}[\bm{r}]\d^3r\right\rangle_t[z_n] \\
 &\simeq\frac{1}{V_z}\sum_{pl\alpha}\mathcal{G}_\Delta[\bm{r}_\alpha]\bm{r}_\alpha\bm{f}^p_lw_{nl}[z_l+r_{z\alpha}]\Delta V^p_l\Delta^3,
\end{split} \label{SigmaF3}
\end{align}
where $V^p_\alpha$ is the volume of particle $p$ contained within the DNS grid cell $\alpha$, the index $l$ ($l_q$) enumerates the grid cells of the Lagrangian layer (of a particle $q$), co-moving with the particles, $\Delta V^p_l$ ($\Delta V^q_{l_q}$) is the volume contained within the grid cell $l$ ($l_q$), and $w_{nl}[z]$ is the fraction of the line connecting $z$ with $z_l$ that is contained within the discretization interval $[z_n-\Delta z/2,z_n+\Delta z/2)$:
\begin{align}
 w_{nl}[z]&=
 \begin{cases}
  0&\text{if}\quad l_{nl}[z]<0 \\
	1&\text{if}\quad l_{nl}[z]=0\land z=z_l \\
	\frac{l_{nl}[z]}{|z-z_l|}&\text{else},
 \end{cases}
\end{align}
with
\begin{equation}
\begin{split}
 l_{nl}[z]\equiv&\;\min\left(\max(z,z_l),z_n+\Delta z/2\right) \\
 &-\max\left(\min(z,z_l),z_n-\Delta z/2\right).
\end{split}
\end{equation}
Note that, in the IBM, the Lagrangian surface grid is usually discretized into very thin volume grid cells of volume $\Delta^3$ and the surface force density $\bm{\tilde f}^p_{\mathrm{IBM}l}$ is therefore represented by a volumetric force density $\bm{f}^p_l$~\citep{Uhlmann05}. These two densities are related to each other through $\bm{\tilde f}^p_{\mathrm{IBM}l}S^p_l=\bm{f}^p_l\Delta^3$.

\subsection{First-order approximations (for comparison only)} \label{ComparisonFry}
Here, we briefly describe the approximations that need to be carried out in order to obtain a coarse-graining formulation equivalent to that of \citet{Fryetal24}, who essentially used the expressions by \citet{Jackson97}, approximated in the first order of $R/L$. To achieve the same level of approximation, the weights $w^p_n[z]$ and $w_{nl}[z]$ in Eqs.~(\ref{SigmaC}) and (\ref{SigmaF1})-(\ref{SigmaF3}), weighing equally along a line connecting $z^p$ and $z^l$, respectively, with $z$, are replaced by the singular weights $\left(w^p_n\right)_1$ and $\left(w_{nl}\right)_1$, respectively:
\begin{align}
 \left(w^p_n\right)_1&\equiv
 \begin{cases}
  1&\text{if}\quad z^p\in[z_n-\Delta z/2,z_n+\Delta z/2)\\
	0&\text{else},
 \end{cases} \\
 \left(w_{nl}\right)_1&\equiv
 \begin{cases}
  1&\text{if}\quad z_l\in[z_n-\Delta z/2,z_n+\Delta z/2)\\
	0&\text{else}.
 \end{cases}
\end{align}
This is because, in the first order of $R/L$, the delta lines in Eqs.~(\ref{SigmaC}) and (\ref{SigmaF1})-(\ref{SigmaF3}) become standard delta distributions localized in $z^p$ and $z_l$, respectively. Note that there are no changes applied to the stresses in Eqs.~(\ref{FhDelta}) and (\ref{FcDelta}), since they exactly compensate the differences between force densities based on the solid volume fraction $\beta_\mathrm{s}$ in our method, such as $\beta_\mathrm{s}\langle\bm{f}_\mathrm{h}\rangle^\mathrm{s}=\langle\sum_pX^p\bm{f}^p_\mathrm{h}\rangle$, and those based on the number density in \citet{Fryetal24}, such as $\langle\sum_p\bm{F}^p_\mathrm{h}\delta[\bm{x}-\bm{x}^p]\rangle$.
\end{appendices}

%\bibliography{model}

\end{document}